\documentclass[10pt,journal]{IEEEtran}

\usepackage{amsmath,epsfig,amssymb,verbatim}
\usepackage{cite}
\usepackage{graphicx}
\usepackage{bm,bbm}
\usepackage{xcolor}
\usepackage{hhline}
\usepackage{subfigure}
\usepackage{stackengine}
\usepackage{cleveref}
\usepackage{multicol}
\usepackage{multirow}
\usepackage{algorithm,algpseudocode}

\def\@IEEEBIOskipN{1\baselineskip}

\newtheorem{Proposition}{Proposition}

\begin{document}
\title{Cross-Band Movable-Antenna Communications under Frequency-Dependent Multipath}

\author{Guangchen Wang, \IEEEmembership{Member, IEEE,} Zhifeng Tang, \IEEEmembership{Member, IEEE,} Haiyao Yu, \IEEEmembership{Member, IEEE,}\\ Nan Yang, \IEEEmembership{Senior Member, IEEE,} Salman Durrani, \IEEEmembership{Senior Member, IEEE,} and Xiangyun Zhou, \IEEEmembership{Fellow, IEEE}
\vspace{-2em}
\thanks{G. Wang, Z. Tang, N. Yang, S. Durrani, and X. Zhou are with the School of Engineering, Australian National University, Canberra, ACT 2601, Australia (Emails: \{guangchen.wang, zhifeng.tang, nan.yang, salman.durrani, xiangyun.zhou\}@anu.edu.au).}
\thanks{H. Yu is with the School of Electrical and Computer Engineering, The University of Sydney, Sydney, NSW 2006, Australia (e-mail: haiyao.yu@sydney.edu.au).}

}

\maketitle

\begin{abstract}
Movable antennas (MAs) provide additional spatial degrees of freedom by adapting their physical positions to the propagation environment. However, existing studies mainly focus on single-band or contiguous wideband transmission, leaving the cross-band spatial behavior of MAs under frequency-dependent multipath propagation largely unexplored. To address this gap, we investigate a downlink multi-band MA communication system where a common physical MA configuration is shared across multiple communication bands. Such sharing induces an inherent configuration conflict, as different bands often favor substantially divergent MA configurations. We first develop a frequency-dependent multipath channel model with shared propagation geometry but band-dependent multipath component powers and phases. Using this model, we formulate a sum-rate maximization problem by jointly optimizing user-band assignment, MA positioning, and transmit precoding. We further establish a cross-band spatial scaling relationship for the normalized channel power and derive both a position-independent channel gain upper bound and a sum-rate upper bound. Based on these results, we propose a cross-band analysis-guided optimization (CAGO) algorithm and a cross-band graph transformer (CBGT) to reduce online optimization complexity. Numerical results validate the derived cross-band spatial relationship and show that CAGO consistently outperforms the narrowband, wideband, and fixed-antenna benchmarks while remaining close to the idealized independent-MA benchmark. Additionally, CBGT with online refinement achieves near-CAGO sum-rate performance with substantially lower online computational cost, providing a practical framework for exploiting the spatial reconfiguration capability of MAs in multi-band communication systems.
\end{abstract}

\vspace{-1em}

\section{Introduction}

The rapid evolution of next-generation wireless networks, particularly sixth-generation (6G) systems, calls for highly flexible and adaptive transmission architectures to support increasingly diverse communication requirements \cite{Zhang2019vtm,Ngo2024iotj,Tang2023JSAC,wang2026distributedoptimizationlearninggraphtransformers}. Conventional fixed-position antenna arrays are inherently limited in their ability to dynamically adapt to channel variations and spatial user distributions, especially in complex propagation environments. To address this limitation, movable antenna (MA) technology has recently emerged as a promising paradigm for wireless communications \cite{Zhu2024MAOverview,Zhu2024MAModeling}. Unlike conventional antenna architectures with fixed spatial sampling locations, MAs can dynamically adjust their physical positions within a predefined region to better match the propagation environment. By enabling spatial reconfiguration of antenna positions, MAs introduce additional spatial degrees of freedom (DoFs), allowing wireless systems to exploit spatial channel variations that are otherwise inaccessible to fixed-position architectures \cite{Ma2024MIMOCapacity}. 

Motivated by these advantages, studies have investigated the fundamental performance limits, position optimization, and practical implementation of MA-enabled wireless systems \cite{Mei2024GraphMA,Ning2025MAArchitecture,Ding2025PinchingPerspective}. Early studies first characterized the position-dependent multipath channel response of MAs, revealing how antenna displacement within a local region alters the phases of different multipath components (MPCs) and thereby induces spatial channel variations \cite{Zhu2024MAModeling}. Building on this channel characterization, subsequent studies optimized the MA positions to exploit such spatial variations and improve the communication performance, initially focusing primarily on relatively simple single-user scenarios. The design becomes substantially more involved in multiuser systems, since the MA positions are coupled with the desired and interfering channels of multiple UEs. To address this coupling, MA positioning has been jointly optimized with transceiver and resource-allocation variables, including transmit precoding, receive combining, and power control, for multiuser communication systems with different MA deployment configurations \cite{Zhu2024MultiuserMA,Yang2024FlexiblePrecoding,Xiao2024MABS}. Beyond conventional multiuser transmission, MA positioning has also been incorporated into more sophisticated multiple-access and cooperative transmission schemes, such as non-orthogonal multiple access (NOMA) and coordinated multipoint transmission \cite{Li2024MANOMA,Hu2024CoMPMA}. In addition, MA positioning has been exploited for physical-layer security, where it can be jointly designed with beamforming, power allocation, and artificial noise to enhance legitimate transmission while suppressing information leakage toward eavesdroppers \cite{Hu2024SecureMA,Tang2025SecureMIMO,Shen2025wcl,tang2026artificialnoiseaideddesignmovableantenna}. These studies have shown that MA positioning needs to be jointly coordinated with transceiver and resource-allocation design to accommodate the competing channel conditions of multiple users and links. However, such studies have assumed narrowband operation within a single frequency band, where the position-dependent channel response is characterized at a single carrier frequency.

Beyond such single-band operation, future wireless systems are expected to exploit heterogeneous spectrum resources distributed across multiple frequency bands to accommodate diverse service requirements and leverage their complementary bandwidth and propagation characteristics \cite{Aboagye2024Multiband,Flinck2025Spectrum}. In particular, 6G multi-band operation may span sub-6 GHz, millimeter-wave, and low-terahertz frequencies, enabling the joint use of widely separated spectrum resources with complementary coverage and capacity characteristics \cite{Katwe2025_6GSpectrum,THzComMCS2024}. This introduces an additional frequency-domain dimension into the design of MA-enabled communication systems. Along this direction, recent studies have extended MA-enabled communications from narrowband to wideband transmission scenarios, where antenna positioning needs to account for channel variations across multiple frequency components. In particular, MA positioning has been investigated for wideband orthogonal frequency-division multiplexing (OFDM) systems under frequency-selective multipath channels, where a common antenna configuration needs to balance the channel conditions across multiple subcarriers \cite{Zhu2024WidebandMA,Irshad2025WidebandMA,Tu2026MAOFDM}. Nevertheless, these studies have mainly considered frequency-selective transmission within a contiguous communication band, where the frequency components remain relatively close to the center frequency of the operating band. Consequently, the resulting spatial channel variations across frequency are relatively limited, and recent results have shown that, under certain OFDM settings, MA positions optimized based on a narrowband channel model can achieve performance close to that obtained with full wideband optimization \cite{Feng2025MISOOFDM}. 

In contrast, multi-band transmission may involve multiple distinct communication bands with substantially different carrier frequencies \cite{Saeidi2024_Multiband}. Compared with the relatively smooth frequency-dependent channel variations within a contiguous wideband system, large carrier-frequency separations in multi-band transmission can lead to substantially more pronounced differences in the spatial channel landscapes across bands. Although recent frequency-position antenna architectures have jointly exploited frequency allocation and antenna positioning \cite{Shen2024FPFA,Shen2026DFPFA}, these studies have relied on simplified terahertz propagation models and have not explicitly characterized the frequency-dependent variations of the MPCs across different communication bands. Importantly, even when different bands share the same underlying propagation geometry, their effective multipath channels can differ substantially due to wavelength-dependent spatial phase responses and frequency-dependent variations in the relative powers and phases of the MPCs \cite{3GPP38901}. As a result, different bands may favor substantially different MA configurations, creating an inherent cross-band MA configuration conflict under a common physical MA layout. To the best of our knowledge, this fundamental cross-band configuration problem has not been investigated in the existing MA literature under frequency-dependent multipath propagation. Understanding and resolving this conflict is essential for fully exploiting the spatial reconfiguration capability of MAs in multi-band communication systems. Moreover, practical MAs have finite movement speeds, such that frequent repositioning among band-specific antenna configurations incurs non-negligible movement delay and reduces the effective transmission time \cite{Wang2026MovementDelay,Hu2026MovementTradeoff}. This practical limitation further necessitates a common physical MA configuration that can balance the distinct spatial channel conditions across multiple communication bands. In multiuser scenarios, this conflict is coupled with user-band assignment and multiuser interference, as the band assignment determines which users share the same frequency resources while the MA positions simultaneously shape their channel conditions, making the joint design of user-band assignment, MA positioning, and transmit precoding highly challenging.

\textit{Our Contributions:} This paper investigates a downlink multi-band MA communication system, in which a common physical MA configuration is shared across distinct communication bands under frequency-dependent multipath propagation. To address the resulting cross-band MA configuration conflict and multiuser coupling, we jointly optimize user-band assignment, MA positioning, and transmit precoding to effectively exploit the available frequency and spatial DoFs. Our main contributions are:

\begin{itemize}

    \item We develop a multi-band MA communication framework that explicitly accounts for frequency-dependent multipath propagation across distinct communication bands. The underlying propagation geometry is shared across bands, while the relative MPC powers and phases vary with frequency, resulting in distinct spatial channel variations under a common MA configuration. Based on this framework, we formulate a sum-rate maximization problem by jointly optimizing user-band assignment, MA positioning, and transmit precoding.

    \item We analytically characterize the cross-band spatial channel characteristics of the considered system. We establish a cross-band spatial scaling relationship for the normalized channel power, derive a position-independent channel-gain upper bound, and further obtain an analytical sum-rate upper bound through a band-level resource allocation problem. These results characterize how frequency-dependent MPC variations affect favorable MA positions across bands and provide tractable guidance for resolving the resulting cross-band MA configuration conflict.

    \item We develop a cross-band analysis-guided optimization (CAGO) algorithm that exploits the analytical results for user-band screening, initialization, and joint optimization of the transmit precoders, MA positions, and band assignment. We further develop a cross-band graph transformer (CBGT) to reduce the online computational complexity through offline learning and lightweight online refinement. Using numerical results, we validate the derived cross-band spatial relationship, show that CAGO consistently outperforms the considered practical benchmarks while remaining close to {an idealized benchmark that removes the common MA-configuration constraint and allows each band to independently optimize its MA configuration}, and that CBGT achieves near-CAGO sum-rate performance with substantially lower online computational cost.
     
\end{itemize}

The rest of the paper is organized as follows. Section~\ref{Sec:System} presents the considered multi-band MA system model and formulates the joint user-band assignment, MA positioning, and transmit precoding problem. Section~\ref{Sec:Anal} analyzes the cross-band spatial channel characteristics and derives analytical performance upper bounds. Section~\ref{Sec:CAGO} develops the proposed CAGO algorithm and CBGT framework for solving the formulated optimization problem. Section~\ref{Sec:Num} presents numerical results to evaluate the proposed schemes and provide further insights into multi-band MA transmission. Finally, Section~\ref{Sec:Con} concludes the paper.

\section{System Model}\label{Sec:System}

\begin{figure}[t!]
    \centering
   \includegraphics[width=0.75\columnwidth]{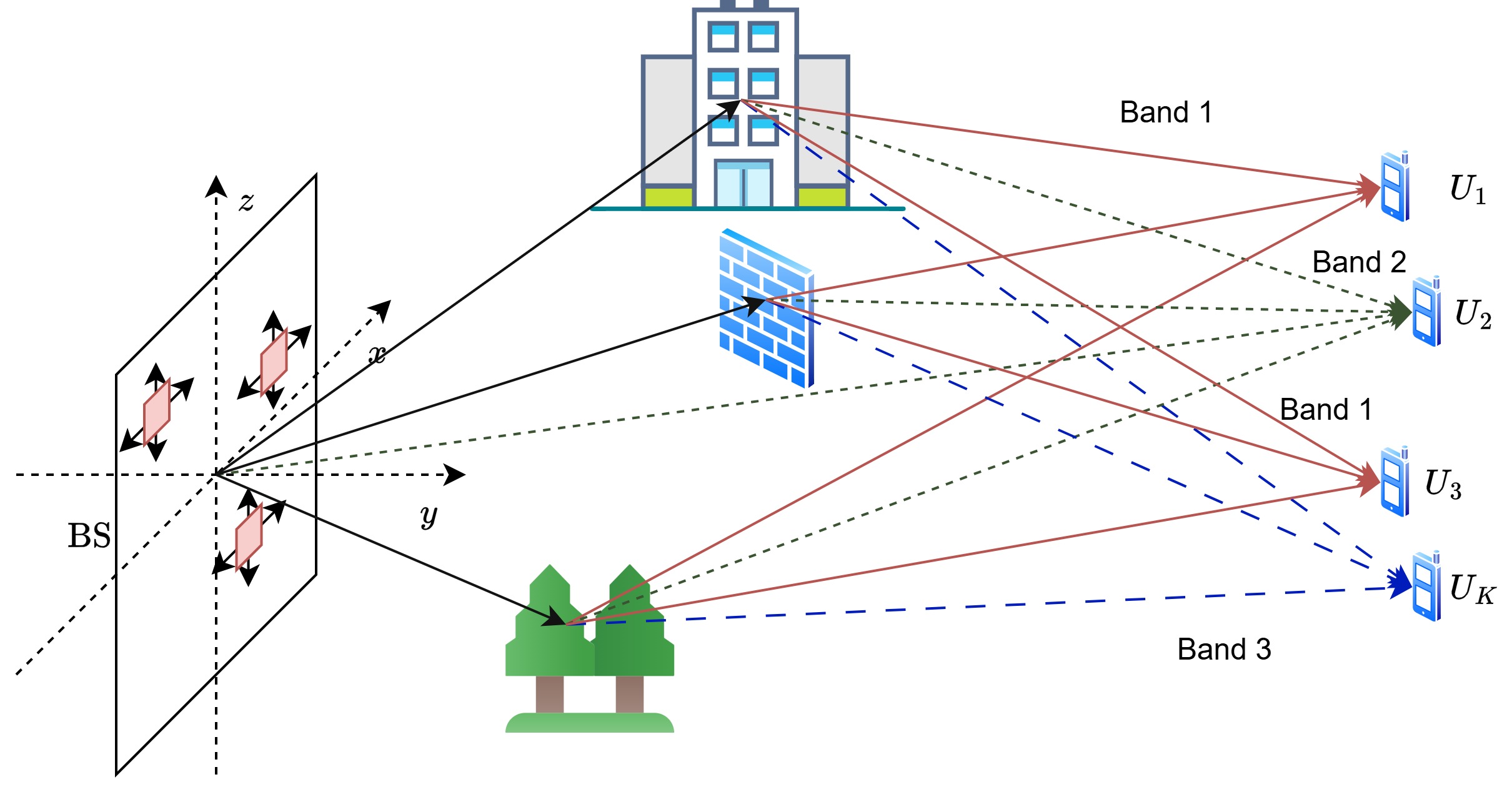}\vspace{-1.5em}
  \caption{Illustration of the considered multi-band MA system.}
    \label{fig:System_Model}
    \vspace{-0.5em}
\end{figure}

We consider a downlink multi-band communication system, where a base station (BS) serves $K$ single-antenna UEs, denoted by $U_k$, $k\in\mathcal{K}\triangleq\{1,\ldots,K\}$, as depicted in Fig.~\ref{fig:System_Model}. The BS supports transmission over $B$ distinct communication bands, indexed by $b\in\mathcal{B}\triangleq\{1,\ldots,B\}$. The carrier frequency and bandwidth of band $b$ are denoted by $f_b$ and $W_b$, respectively. The considered bands occupy orthogonal frequency resources and therefore do not cause inter-band interference.

The BS is equipped with an MA platform supporting $M$ MAs. We establish a three-dimensional (3D) Cartesian coordinate system with the BS located at the origin. Without loss of generality, the MA aperture is assumed to lie on the $x$-$z$ plane, with the $y$-axis perpendicular to the aperture surface. Accordingly, the position of the $m$-th MA is denoted by $\mathbf{q}_m=[x_m,0,z_m]^T$, where $x_m$ and $z_m$ denote its coordinates within the aperture. The positions of all MAs are collected in the matrix $\mathbf{Q}=[\mathbf{q}_1,\ldots,\mathbf{q}_M]\in\mathbb{R}^{3\times M}$. We assume that each MA can move within a predefined aperture region $\mathcal{Q}$, i.e., $\mathbf{q}_m\in\mathcal{Q}$, $\forall m$. To avoid antenna collision and excessive electromagnetic coupling, a minimum spacing constraint between any two MAs is imposed as
\begin{equation}\label{eq:MA_spacing}
    \|\mathbf{q}_i-\mathbf{q}_j\|\geq d_{\min}, \quad \forall i\neq j,
\end{equation}
where $d_{\min}$ denotes the minimum inter-MA distance and $\|\cdot\|$ is the Euclidean norm. Importantly, the same physical MA configuration $\mathbf{Q}$ is shared by all communication bands, while the transmit precoding can be independently designed for different bands.

The position of $U_k$ is denoted by $\mathbf{p}_k=[x_{U,k},y_{U,k},z_{U,k}]^T$. We define $\mathbf{q}_0=[0,0,0]^T$ as a reference position on the MA platform. Accordingly, the distance between $U_k$ and $\mathbf{q}_0$ is
given by
\begin{equation}\label{eq:UE_distance}
    d_k=\|\mathbf{p}_k-\mathbf{q}_0\|.
\end{equation}
We assume that the UE locations remain fixed during each transmission block. Multiple UEs may be served over the same communication band and spatially multiplexed through the MAs at the BS.

\subsection{Multi-Band Multipath Channel Model}

We adopt a geometric multipath channel model, where the BS--$U_k$ link consists of $L_k$ effective MPCs. Since the considered communication bands propagate through the same physical environment, the underlying propagation geometry is assumed to be shared across the bands. In particular, the propagation direction of each MPC remains unchanged across the considered bands, while its effective power may vary with the operating frequency. Under the plane-wave approximation, the channel coefficient between the $m$-th MA and $U_k$ over band $b$ is expressed as
\begin{equation}\label{eq:channel_coefficient}
    h_{k,m}^{(b)}(\mathbf{q}_m)=\sum_{\ell=1}^{L_k}\alpha_{k,\ell}^{(b)}\exp\left(-j\frac{2\pi}{\lambda_b}\mathbf{u}_{k,\ell}^{T}(\mathbf{q}_m-\mathbf{q}_0)\right),
\end{equation}
where $\alpha_{k,\ell}^{(b)}$ denotes the complex gain of the $\ell$-th MPC over band $b$ at the reference position $\mathbf{q}_0$, and $\mathbf{u}_{k,\ell}\in\mathbb{R}^{3\times1}$ denotes the corresponding unit propagation direction vector. Specifically, $\mathbf{u}_{k,\ell}$ is given by
\begin{equation} \label{eq:direction_vector}
    \mathbf{u}_{k,\ell}=[\sin\theta_{k,\ell}\cos\phi_{k,\ell}, \cos\theta_{k,\ell}, \sin\theta_{k,\ell}\sin\phi_{k,\ell}]^T,
\end{equation}
where $\theta_{k,\ell}$ and $\phi_{k,\ell}$ denote the polar angle measured from the positive $y$-axis and the azimuth angle of the $\ell$-th MPC, respectively.

To characterize the complex gain of each MPC, we express $\alpha_{k,\ell}^{(b)}$ in terms of the large-scale channel gain, relative MPC power, and reference phase as
\begin{equation}\label{eq:path_gain}
\alpha_{k,\ell}^{(b)}=\sqrt{\beta_k^{(b)}P_{k,\ell}^{(b)}}e^{j\varphi_{k,\ell}^{(b)}},
\end{equation}
where $\beta_k^{(b)}$ denotes the large-scale channel gain between the BS and $U_k$ over band $b$, $P_{k,\ell}^{(b)}$ denotes the normalized power of the $\ell$-th MPC, and $\varphi_{k,\ell}^{(b)}$ denotes the phase of the $\ell$-th MPC at the reference position $\mathbf{q}_0$ over band $b$. The reference phase $\varphi_{k,\ell}^{(b)}$ incorporates the propagation-delay-induced phase and the frequency-dependent interaction phase, and therefore generally varies across communication bands \cite{3GPP38901}. The physical propagation geometry, including the MPC directions and propagation delays, is assumed to be shared across the considered communication bands, while the complex path gains vary with frequency due to the frequency-dependent attenuation, relative MPC powers, and reference phases. The large-scale channel gain is given by $\beta_k^{(b)}=\left(\frac{c}{4\pi f_b}\right)^2d_k^{-\eta}$, where $\eta$ denotes the path-loss exponent.

{Following the multi-frequency channel modeling principle in \cite{3GPP38901}, variations in the relative MPC powers across communication bands are modeled through band-dependent per-cluster shadowing.} Such variations arise because different MPCs interact with different propagation objects and surfaces, whose reflection, scattering, and penetration characteristics are frequency dependent, leading to distinct attenuation levels. Accordingly, the unnormalized power of the $\ell$-th MPC over band $b$ is expressed as
\begin{equation}\label{eq:unnormalized_path_power}
    \widetilde{P}_{k,\ell}^{(b)} =10^{-Z_{k,\ell}^{(b)}/10},
\end{equation}
where $Z_{k,\ell}^{(b)}$ denotes the corresponding band-dependent per-cluster shadowing term, modeled as an independent Gaussian random variable across communication bands and MPCs, i.e.,  $Z_{k,\ell}^{(b)}\sim\mathcal{N}(0,\zeta^2)$, with $\zeta$ denoting the per-cluster shadowing standard deviation in dB. The normalized MPC power is then given by
\begin{equation}\label{eq:normalized_path_power}
    P_{k,\ell}^{(b)}=\frac{\widetilde{P}_{k,\ell}^{(b)}}{\sum_{r=1}^{L_k}\widetilde{P}_{k,r}^{(b)}},
\end{equation}
which satisfies $\sum_{\ell=1}^{L_k}P_{k,\ell}^{(b)}=1$. Although $Z_{k,\ell}^{(b)}$ is statistically modeled as a random variable across channel realizations, its realization remains fixed within each channel realization. We assume that perfect channel state information (CSI) over all considered communication bands is available at the BS prior to the user-band assignment, MA positioning, and transmit precoding design. Accordingly, the realizations of the multipath parameters, including the propagation directions, complex path gains, and band-dependent MPC powers, are assumed to be perfectly known at the BS. {Since all MAs are located within the same antenna aperture, they share the same underlying multipath geometry and path gains for a given UE and communication band, while their distinct positions induce different path-dependent phase shifts.} By stacking the channel coefficients associated with all $M$ MAs, the channel vector between the BS and $U_k$ over band $b$ is expressed as
\begin{equation}\label{eq:channel_vector}
    \mathbf{h}_k^{(b)}(\mathbf{Q})=\left[h_{k,1}^{(b)}(\mathbf{q}_1),\ldots,h_{k,M}^{(b)}(\mathbf{q}_M)\right]^T\in\mathbb{C}^{M\times1}.
\end{equation}

Although the underlying propagation geometry is shared across the considered communication bands, their spatial channel variations with respect to the MA positions are generally different. This difference arises from both the wavelength-dependent spatial phase in \eqref{eq:channel_coefficient} and the band-dependent relative MPC powers in \eqref{eq:normalized_path_power}. Consequently, an MA configuration that provides favorable channel conditions over one band may not necessarily be favorable over another band.

\subsection{Multi-Band Transmission}

Each UE is assigned to one of the $B$ communication bands for downlink transmission. We introduce a binary band-assignment variable $a_{k,b}\in\{0,1\}$, where $a_{k,b}=1$ indicates that $U_k$ is assigned to band $b$, and $a_{k,b}=0$ otherwise. Each UE is assigned to exactly one communication band, i.e., $\sum_{b=1}^{B}a_{k,b}=1,\quad \forall k\in\mathcal{K}$. Accordingly, the set of UEs assigned to band $b$ is defined as $\mathcal{K}_b=\left\{k\in\mathcal{K}:a_{k,b}=1\right\}$.

The BS employs linear precoding to spatially multiplex the UEs assigned to the same communication band. The transmitted signal over band $b$ is expressed as
\begin{equation}\label{eq:transmit_signal}
    \mathbf{x}_b
    =
    \sum_{k\in\mathcal{K}_b}
    \mathbf{w}_{k,b}s_{k,b},
\end{equation}
where $\mathbf{w}_{k,b}\in\mathbb{C}^{M\times1}$ denotes the precoding vector for $U_k$ over band $b$, and $s_{k,b}$ denotes the corresponding information symbol satisfying $\mathbb{E}[|s_{k,b}|^2]=1$. For $U_k$ assigned to band $b$, the received signal is given by
\begin{equation}\label{eq:received_signal}
    y_{k,b}\!=\!\mathbf{h}_k^{(b)H}(\!\mathbf{Q}\!)\mathbf{w}_{k,b}s_{k,b}\!+\!\!\!\sum_{j\!\in\!\mathcal{K}_b,\substack{j\!\neq\! k}}\!\mathbf{h}_k^{(b)H}(\!\mathbf{Q}\!)\mathbf{w}_{j,b}s_{j,b}\!\!+\!n_{k,b},
\end{equation}
where $n_{k,b}\sim\mathcal{CN}(0,\sigma_b^2)$ denotes the additive white Gaussian noise (AWGN) over band $b$. Since the considered communication bands are orthogonal in frequency, $U_k$ only experiences multiuser interference from the other UEs assigned to the same band. Accordingly, the signal-to-interference-plus-noise ratio (SINR) of $U_k$ over band $b$ is expressed as
\begin{equation}\label{eq:SINR}
    \gamma_{k,b}=\frac{\left\| \mathbf{h}_k^{(b)H}(\mathbf{Q})\mathbf{w}_{k,b}\right\|^2}{\displaystyle\sum_{\substack{j\in\mathcal{K}_b,j\neq k}}\left\|\mathbf{h}_k^{(b)H}(\mathbf{Q}) \mathbf{w}_{j,b} \right\|^2+\sigma_b^2}.
\end{equation}
The corresponding achievable rate is then given by
\begin{equation}\label{eq:achievable_rate}
    R_{k,b}=W_b\log_2\left(1+\gamma_{k,b}\right).
\end{equation}
Therefore, the achievable rate of $U_k$ under the band assignment $\{a_{k,b}\}$ is expressed as
\begin{equation}\label{eq:user_rate}
    R_k=\sum_{b=1}^{B}a_{k,b}R_{k,b}.
\end{equation}

\subsection{Problem Formulation}

In the considered multi-band MA system, our design aims to maximize the sum rate of all UEs by jointly optimizing the user-band assignment, MA positions, and transmit precoding. We collect the band-assignment variables into $\mathbf{A}=[a_{k,b}]\in\{0,1\}^{K\times B}$ and denote the collection of transmit precoders by $\mathbf{W}\triangleq \{\mathbf{w}_{k,b}:k\in\mathcal{K},\,b\in\mathcal{B}\}$. Accordingly, the optimization problem is formulated as
\begin{subequations}\label{eq:P1}
\begin{align}
    \mathrm{P1}:~\max_{\mathbf{A},\mathbf{Q},\mathbf{W}} ~~&\sum_{k=1}^{K} R_k\label{eq:P1_obj}\\
    \mathrm{s.t.}~~&\sum_{b=1}^{B}a_{k,b}=1,\quad \forall k\in\mathcal{K},\label{eq:P1_assignment}\\
    &a_{k,b}\in\{0,1\},\quad \forall k\in\mathcal{K},~b\in\mathcal{B},\label{eq:P1_binary}\\
    &\sum_{b=1}^{B}a_{k,b}\gamma_{k,b}\geq \gamma_k^{\mathrm{th}},\quad \forall k\in\mathcal{K}, \label{eq:P1_SINR}\\
    &\sum_{b=1}^{B}\sum_{k=1}^{K}a_{k,b}\|\mathbf{w}_{k,b}\|^2\leq P_{\max},\label{eq:P1_power}\\
    &\mathbf{q}_m\in\mathcal{Q},\quad \forall m,\label{eq:P1_MA_region}\\
    &\|\mathbf{q}_i-\mathbf{q}_j\|\geq d_{\min},\quad \forall i\neq j,\label{eq:P1_MA_spacing}
\end{align}
\end{subequations}
where $\gamma_k^{\mathrm{th}}$ denotes the UE-specific minimum SINR requirement of $U_k$, which may vary across UEs to accommodate heterogeneous QoS requirements, and $P_{\max}$ denotes the maximum transmit power available at the BS. Constraints \eqref{eq:P1_assignment} and \eqref{eq:P1_binary} ensure that each UE is assigned to exactly one communication band. Constraint \eqref{eq:P1_SINR} guarantees the minimum SINR requirement of each UE over its assigned band, while constraint \eqref{eq:P1_power} limits the total transmit power across all communication bands. Constraints \eqref{eq:P1_MA_region} and \eqref{eq:P1_MA_spacing} restrict the MAs to the predefined aperture and guarantee the minimum inter-MA spacing, respectively.

Problem P1 is a mixed-integer non-convex optimization problem due to the binary band-assignment variables and the coupling among the user-band assignment, MA positions, and transmit precoding. Specifically, the band assignment determines which UEs share the same frequency resources and thus interfere with each other, while the common MA configuration $\mathbf{Q}$ simultaneously affects the spatial channels over all communication bands. Moreover, due to the wavelength-dependent spatial phase variations and band-dependent relative MPC powers, different communication bands generally exhibit different spatial channel variations with respect to the MA positions. Therefore, the user-band assignment, MA positions, and transmit precoding need to be jointly designed to effectively exploit the available frequency and spatial degrees of freedom.

\section{Cross-Band Channel and Performance Analysis}\label{Sec:Anal}

\subsection{Cross-Band Spatial Channel Characteristics}

To characterize the position-dependent channel variations across different communication bands, we first focus on the channel between a single MA and $U_k$. To isolate the spatial channel variation from the band-dependent large-scale channel gain, we define the normalized channel power over band $b$ as
\begin{equation}\label{eq:normalized_spatial_gain}
    G_k^{(b)}(\mathbf{q})=\frac{|h_k^{(b)}(\mathbf{q})|^2}{\beta_k^{(b)}}.
\end{equation}
The following proposition establishes a fundamental cross-band relationship between the normalized spatial channel landscapes.

\begin{Proposition}\label{Proposition:1}
For two communication bands $b_1$ and $b_2$, if their normalized MPC power profiles and pairwise MPC phase differences are identical, i.e., $P_{k,\ell}^{(b_1)}=P_{k,\ell}^{(b_2)}$ and $\varphi_{k,\ell}^{(b_1)} -\varphi_{k,r}^{(b_1)}=\varphi_{k,\ell}^{(b_2)}-\varphi_{k,r}^{(b_2)}$, $\forall \ell,r \in \{1,\ldots,L_k\}$, their normalized spatial channel powers satisfy
\begin{equation}\label{eq:cross_band_scaling}
    G_k^{(b_2)}(\mathbf{q}_0+\mathbf{d})=G_k^{(b_1)}\left(\mathbf{q}_0+\frac{f_{b_2}}{f_{b_1}}\mathbf{d}\right),
\end{equation}
for any spatial displacement $\mathbf{d}$. 

\begin{IEEEproof}
Based on \eqref{eq:channel_coefficient} and \eqref{eq:path_gain}, the normalized channel power in \eqref{eq:normalized_spatial_gain} is expressed as
\begin{align}\label{eq:spatial_gain_expansion}
    G_k^{(b)}(\mathbf{q})&=1+2\sum_{\ell<r}\sqrt{P_{k,\ell}^{(b)}P_{k,r}^{(b)}}\notag\\
    &\quad\times\cos\left(\Delta\varphi_{k,\ell r}^{(b)}-\frac{2\pi}{\lambda_b}\Delta\mathbf{u}_{k,\ell r}^{T}(\mathbf{q}-\mathbf{q}_0)\right),
\end{align}
where $\Delta\varphi_{k,\ell r}^{(b)}=\varphi_{k,\ell}^{(b)}-\varphi_{k,r}^{(b)}$ and $\Delta\mathbf{u}_{k,\ell r}=\mathbf{u}_{k,\ell}-\mathbf{u}_{k,r}$. For band $b_1$, evaluating \eqref{eq:spatial_gain_expansion} at $\mathbf{q}=\mathbf{q}_0+\frac{f_{b_2}}{f_{b_1}}\mathbf{d}$ yields the spatial phase term as
\begin{equation}
    \frac{2\pi}{\lambda_{b_1}}
    \Delta\mathbf{u}_{k,\ell r}^{T}
    \frac{f_{b_2}}{f_{b_1}}\mathbf{d}
    =
    \frac{2\pi}{\lambda_{b_2}}
    \Delta\mathbf{u}_{k,\ell r}^{T}\mathbf{d},
\end{equation}
where $\lambda_b=c/f_b$ is used. Since the normalized MPC powers and pairwise MPC phase differences are identical across bands $b_1$ and $b_2$, each term in \eqref{eq:spatial_gain_expansion} for band $b_1$ evaluated at $\mathbf{q}_0+\frac{f_{b_2}}{f_{b_1}}\mathbf{d}$ is identical to the corresponding term for band $b_2$ evaluated at $\mathbf{q}_0+\mathbf{d}$, which yields \eqref{eq:cross_band_scaling}.
\end{IEEEproof}
\end{Proposition}

\textit{Remark 1:} Proposition~\ref{Proposition:1} shows that, when the normalized MPC powers and pairwise MPC phase differences remain unchanged across two communication bands, their normalized spatial channel landscapes preserve the same structure up to a frequency-dependent spatial scaling. In practical multi-band channels, however, the normalized MPC powers and MPC phases generally vary with frequency \cite{3GPP38901}. As a result, variations in the pairwise MPC phase differences shift the locations of constructive and destructive multipath combinations over the MA aperture, while variations in the normalized MPC powers alter the contributions of different MPCs to the resulting channel gain. Consequently, different communication bands may exhibit different favorable and unfavorable MA locations. The following proposition further derives a position-independent upper bound on the channel gain.

\begin{Proposition}\label{Proposition:2}
For any MA position $\mathbf{q}\in\mathcal{Q}$, the channel power between a single MA and $U_k$ over band $b$ is upper bounded by
\begin{equation}\label{eq:single_MA_channel_UB}
    |h_k^{(b)}(\mathbf{q})|^2\leq\beta_k^{(b)}\left(\sum_{\ell=1}^{L_k}\sqrt{P_{k,\ell}^{(b)}}\right)^2.
\end{equation}
Accordingly, the channel-vector gain with $M$ MAs satisfies
\begin{equation}\label{eq:channel_vector_UB}
    \left\|\mathbf{h}_k^{(b)}(\mathbf{Q})\right\|^2\leq M\beta_k^{(b)}\left(\sum_{\ell=1}^{L_k}\sqrt{P_{k,\ell}^{(b)}}\right)^2.
\end{equation}
\begin{IEEEproof}
Based on \eqref{eq:spatial_gain_expansion} and $\cos(\cdot)\leq 1$, the normalized channel power is upper bounded by
\begin{align}
    G_k^{(b)}(\mathbf{q})\leq1+2\sum_{\ell<r}\sqrt{P_{k,\ell}^{(b)}P_{k,r}^{(b)}} =\left(\sum_{\ell=1}^{L_k}\sqrt{P_{k,\ell}^{(b)}}\right)^2.
\end{align}
Multiplying both sides by $\beta_k^{(b)}$ yields \eqref{eq:single_MA_channel_UB}. Moreover, since
$\left\|\mathbf{h}_k^{(b)}(\mathbf{Q})\right\|^2=\sum_{m=1}^{M}|h_{k,m}^{(b)}(\mathbf{q}_m)|^2$, applying \eqref{eq:single_MA_channel_UB} to each MA yields \eqref{eq:channel_vector_UB}.
\end{IEEEproof}
\end{Proposition}

Proposition~\ref{Proposition:2} provides a position-independent upper bound on the channel gain by assuming ideal coherent combining among all MPCs. Equality in \eqref{eq:single_MA_channel_UB} requires all MPCs to be simultaneously phase-aligned at the same MA position. In general, such simultaneous phase alignment may not be achievable, particularly in rich multipath environments with a large number of MPCs, since the MA position provides only limited spatial DoFs and is further constrained by the finite aperture region. Therefore, \eqref{eq:single_MA_channel_UB} serves as a theoretical upper bound on the channel gain that is not necessarily attainable in general. Nevertheless, the derived bound eliminates the dependence on the MA position and explicitly captures the effect of the normalized MPC power profile, thereby providing a tractable basis for deriving a system-level performance upper bound in the following subsection.

\subsection{Analytical Sum-Rate Upper Bound}

Based on the channel-gain bound in Proposition~\ref{Proposition:2}, we now derive an analytical upper bound on the optimal sum rate of the considered multi-band MA system. For notational simplicity, we define $\overline{g}_{k,b} =M\beta_k^{(b)}\left(\sum_{\ell=1}^{L_k}\sqrt{P_{k,\ell}^{(b)}}\right)^2$ and $\widehat{g}_b=\max_{k\in\mathcal{K}}\overline{g}_{k,b}$. For a given user-band assignment, the effective spatial dimension over band $b$ is upper bounded by $r_b^{\rm o}=\min\{M,|\mathcal{K}_b|\}$, with $0\leq r_b^{\rm o}\leq M$ and $\sum_{b=1}^{B}r_b^{\rm o}\leq K$. To obtain a tractable upper bound, $\{r_b^{\rm o}\}$ is relaxed into continuous effective spatial-dimension variables $\{r_b\}$ satisfying $0\leq r_b\leq M$ and $\sum_{b=1}^{B}r_b\leq K$, while $p_b$ represents the transmit power allocated to communication band $b$.

\begin{Proposition}\label{Proposition:3}
The optimal sum rate $R_{\rm sum}^{\star}$ of problem P1 satisfies
\begin{subequations}\label{eq:sum_rate_UB}
\begin{align}
R_{\rm sum}^{\star}\leq &\overline{R}_{\rm sum}\triangleq\max_{\{p_b\},\{r_b\}}\sum_{b=1}^{B}W_b r_b\log_2\left(1+\frac{p_b\widehat{g}_b}{r_b\sigma_b^2}\right)\label{eq:sum_rate_UB_obj}\\
\mathrm{s.t.}\quad& \sum_{b=1}^{B}p_b\leq P_{\max}, \quad p_b\geq0,\quad \forall b,\label{eq:sum_rate_UB_power1}\\
& \sum_{b=1}^{B}r_b\leq K, \quad 0\leq r_b\leq M,\quad \forall b.\label{eq:sum_rate_UB_dim1}
\end{align}
\end{subequations}
The corresponding term in \eqref{eq:sum_rate_UB_obj} is defined as zero when $r_b\!=\!0$.
\begin{IEEEproof}
For a given user-band assignment and MA configuration, the achievable sum rate with linear precoding over band $b$ is upper bounded by the sum capacity of the corresponding Gaussian broadcast channel. Based on the broadcast-channel and multiple-access-channel duality, the sum capacity is expressed as
\begin{equation}\label{eq:DPC_capacity}
    C_b^{\mathrm{DPC}}(\mathbf{Q})=\max_{\{p_{k,b}\}}W_b\log_2\det\left(\mathbf{I}_M+\frac{1}{\sigma_b^2}\mathbf{S}_b\right),
\end{equation}
subject to $p_{k,b}\geq0$ and $\sum_{k\in\mathcal{K}_b}p_{k,b}\leq p_b$, where $\mathbf{S}_b=\sum_{k\in\mathcal{K}_b}p_{k,b}\mathbf{h}_k^{(b)}(\mathbf{Q})\mathbf{h}_k^{(b)H}(\mathbf{Q})$. The rank of $\mathbf{S}_b$ is upper bounded by $r_b^{\mathrm{o}}=\min\{M,|\mathcal{K}_b|\}$. Based on the determinant--trace inequality, we obtain
\begin{align}\label{eq:DPC_trace_UB}
    C_b^{\mathrm{DPC}}(\mathbf{Q})\!\leq\! W_b r_b^{\mathrm{o}}\log_2\left(1\!+\!\frac{\sum\limits_{k\in\mathcal{K}_b}p_{k,b}\|\mathbf{h}_k^{(b)}(\mathbf{Q})\|^2}{r_b^{\mathrm{o}}\sigma_b^2}\right).
\end{align}
Proposition~\ref{Proposition:2} gives $\|\mathbf{h}_k^{(b)}(\mathbf{Q})\|^2\leq \overline{g}_{k,b}\leq\widehat{g}_b$, and \eqref{eq:DPC_trace_UB} can be further upper bounded as
\begin{equation}\label{eq:band_rate_UB}
    C_b^{\mathrm{DPC}}(\mathbf{Q})\leq W_b r_b^{\mathrm{o}} \log_2\left(1+\frac{p_b\widehat{g}_b}{r_b^{\mathrm{o}}\sigma_b^2}\right).
\end{equation}
The relaxation of $\{r_b^{\mathrm{o}}\}$ into continuous effective spatial-dimension variables $\{r_b\}$ enlarges the feasible set. Optimizing the resulting band-level upper bound over $\{p_b\}$ and $\{r_b\}$ yields \eqref{eq:sum_rate_UB}.
\end{IEEEproof}
\end{Proposition}

The optimization problem in \eqref{eq:sum_rate_UB} is convex since each term in the objective is the perspective of a concave logarithmic function. Therefore, its globally optimal solution can be characterized by the Karush--Kuhn--Tucker (KKT) conditions. The Lagrange multipliers associated with the total transmit-power and spatial-dimension constraints are denoted by $\nu\geq0$ and $\mu\geq0$, respectively. For a given spatial-dimension allocation, the optimal transmit power over band $b$ is obtained as
\begin{equation}\label{eq:UB_power_KKT}
    p_b^{\star}=r_b^{\star}\left[\frac{W_b}{\nu^{\star}\ln 2}-\frac{\sigma_b^2}{\widehat{g}_b}\right]^+.
\end{equation}
The optimal power allocation in \eqref{eq:UB_power_KKT} follows a water-filling structure, where communication bands with larger bandwidths or channel-gain bounds are allocated more transmit power for a given spatial-dimension allocation.

For $r_b^{\star}>0$, the marginal utility with respect to the effective spatial-dimension allocation is expressed as
\begin{equation}\label{eq:UB_DoF_marginal}
\begin{aligned}
    \mathcal{D}_b=\frac{W_b}{\ln 2}\Bigg[\ln\left(1+\frac{p_b^{\star}\widehat{g}_b}{r_b^{\star}\sigma_b^2}\right)-\frac{p_b^{\star}\widehat{g}_b}{r_b^{\star}\sigma_b^2+p_b^{\star}\widehat{g}_b}\Bigg].
\end{aligned}
\end{equation}
Accordingly, the KKT conditions for the effective spatial-dimension variables are given by
\begin{equation}\label{eq:UB_DoF_KKT}
    \begin{cases}
        \mathcal{D}_b\leq\mu^{\star},
        & r_b^{\star}=0,\\
        \mathcal{D}_b=\mu^{\star},
        & 0<r_b^{\star}<M,\\
        \mathcal{D}_b\geq\mu^{\star},
        & r_b^{\star}=M.
    \end{cases}
\end{equation}
where $\mathcal{D}_b$ at $r_b^{\star}=0$ is interpreted as its right-hand limit. The complementary slackness conditions are $\nu^{\star}\left(P_{\max}-\sum_{b=1}^{B}p_b^{\star}\right)=0$ and $\mu^{\star}\left(K-\sum_{b=1}^{B}r_b^{\star}\right)=0$. Substituting \eqref{eq:UB_power_KKT} into \eqref{eq:sum_rate_UB}, the analytical upper bound is equivalently expressed as
\begin{equation}\label{eq:KKT_sum_rate_UB}
    \overline{R}_{\mathrm{sum}}=\sum_{b=1}^{B}W_b r_b^{\star}\left[\log_2\left(\frac{W_b\widehat{g}_b}{\nu^{\star}\ln 2\,\sigma_b^2}\right)\right]^+.
\end{equation}
Equations \eqref{eq:UB_power_KKT}--\eqref{eq:KKT_sum_rate_UB} show that the analytical upper bound can be obtained without enumerating all possible user-band assignments. The relaxed formulation characterizes the band-level allocation of transmit power and effective spatial dimensions according to the bandwidth and channel-gain bound of each communication band, while accounting for the finite total spatial DoFs.

\textit{Remark 2:} The upper bound in Proposition~\ref{Proposition:3} may be loose due to the successive relaxations introduced in its derivation, including the ideal coherent-combining channel-gain bound in Proposition~\ref{Proposition:2}, the Gaussian broadcast-channel sum-capacity bound on linear precoding, the removal of the per-user SINR constraints, and the continuous relaxation of the discrete user-band assignment. Consequently, the derived result provides a tractable analytical upper bound rather than a tight characterization of the achievable sum rate.

\section{Cross-Band Analysis-Guided Optimization}\label{Sec:CAGO}

The analytical results in Section~\ref{Sec:Anal} provide useful guidance for solving problem P1. {Accordingly, we develop a CAGO algorithm that exploits the derived channel-gain and sum-rate bounds in Proposition~\ref{Proposition:2} and \ref{Proposition:3} to reduce the discrete search space and provide effective initialization for the subsequent joint optimization.} Specifically, the channel-gain bound is first used to screen infeasible user-band pairs, while the relaxed band-level solution guides the initialization of the user-band assignment and MA configuration. For a fixed assignment, the transmit precoders and MA positions are alternately updated using weighted minimum mean squared error (WMMSE) and majorization-minimization (MM), respectively. The user-band assignment is subsequently refined through an analysis-guided neighborhood search.

\subsection{Analysis-Guided Screening and Initialization}
\label{Sec:CAGO_initialization}

The position-independent channel-gain bound in Proposition~\ref{Proposition:2} provides a direct criterion for screening infeasible user-band pairs. Based on the previously defined $\overline{g}_{k,b}$, an upper bound on the achievable SINR of $U_k$ over band $b$ is expressed as
\begin{equation}\label{eq:CAGO_screening}
    \overline{\gamma}_{k,b}\triangleq\frac{P_{\max}\overline{g}_{k,b}}{\sigma_b^2},
\end{equation}
where all transmit power is allocated to $U_k$ and multiuser interference
is absent. Accordingly, the candidate-band set of $U_k$ is defined as
\begin{equation}\label{eq:CAGO_candidate_band_set}
    \mathcal{B}_k^{\mathrm{cand}} \triangleq\left\{b\in\mathcal{B}:\overline{\gamma}_{k,b}\geq\gamma_k^{\mathrm{th}} \right\}.
\end{equation}
For any $b\notin\mathcal{B}_k^{\mathrm{cand}}$, the SINR requirement of $U_k$ cannot be satisfied even under the above interference-free upper-bound condition. Hence, the corresponding user-band pair can be safely excluded by fixing $a_{k,b}=0$. If $\mathcal{B}_k^{\mathrm{cand}}=\varnothing$ for any $k$, problem P1 is infeasible.

The relaxed upper-bound problem in \eqref{eq:sum_rate_UB} provides band-level guidance for the initialization of problem P1. Its optimal solution is denoted by $\{p_b^{\mathrm{UB}},r_b^{\mathrm{UB}}\}_{b=1}^{B}$. Although these variables do not directly correspond to a feasible resource allocation for P1, they characterize the preferred band-level allocation of transmit power and effective spatial dimensions in the relaxed system. Based on this solution, the user-band suitability score and the upper-bound-guided target band load are defined as
\begin{equation}\label{eq:CAGO_assignment_score}
    S_{k,b}^{\mathrm{UB}}\triangleq    W_b\log_2\left(1+\frac{p_b^{\mathrm{UB}}\overline{g}_{k,b}}{r_b^{\mathrm{UB}}\sigma_b^2}\right),
\end{equation}
and
\begin{equation}\label{eq:CAGO_target_load}
    \widetilde{n}_b\triangleq\operatorname{Round}_{K}\left(\frac{K r_b^{\mathrm{UB}}}{\sum_{j=1}^{B}r_j^{\mathrm{UB}}}\right),
\end{equation}
respectively, where $\operatorname{Round}_{K}(\cdot)$ denotes a joint rounding operator that maps the continuous band-load values into nonnegative integers satisfying $\sum_{b=1}^{B}\widetilde{n}_b=K$. The suitability score $S_{k,b}^{\mathrm{UB}}$ is set to zero when $r_b^{\mathrm{UB}}=0$, while $\widetilde{n}_b$ represents the target number of UEs assigned to band $b$ according to the relaxed effective spatial-dimension allocation. Based on these quantities, the initial user-band assignment is obtained by solving a reduced mixed-integer linear program, given by
\begin{subequations}\label{eq:CAGO_initial_assignment}
\begin{align}
    \max_{\mathbf{A},\{\xi_b\}}~~&\sum_{k=1}^{K}\sum_{b\in\mathcal{B}_k^{\mathrm{cand}}}a_{k,b}S_{k,b}^{\mathrm{UB}}-\eta_0\sum_{b=1}^{B}\xi_b\label{eq:CAGO_initial_assignment_obj}\\
    \mathrm{s.t.}~~&\sum_{b\in\mathcal{B}_k^{\mathrm{cand}}}a_{k,b}=1,\quad \forall k,\label{eq:CAGO_initial_assignment_unique}\\
    &-\xi_b\leq\sum_{k=1}^{K}a_{k,b}-\widetilde{n}_b\leq\xi_b,\quad \xi_b\geq0, \quad \forall b,\label{eq:CAGO_initial_assignment_load}\\
    &a_{k,b}\in\{0,1\},\quad \forall k,\ b\in\mathcal{B}_k^{\mathrm{cand}}.\label{eq:CAGO_initial_assignment_binary}
\end{align}
\end{subequations}
The auxiliary variable $\xi_b$ characterizes the deviation of the number of
UEs assigned to band $b$ from the target load $\widetilde{n}_b$, while
$\eta_0$ controls the tradeoff between the user-band suitability and the
deviation from the target band loads.

The analytical results are further exploited to initialize the common MA configuration. The communication band contributing the most to the analytical upper bound is selected as the anchor band, denoted by $b_0$. Starting from a uniformly spaced feasible MA configuration, several single-band MA positioning updates are performed over band $b_0$ to obtain the anchor configuration $\mathbf{Q}_{b_0}^{\mathrm{anc}}\in\mathbb{R}^{3\times M}$. Motivated by the cross-band spatial scaling relationship in Proposition~\ref{Proposition:1}, the anchor configuration is further mapped to band $b$ as
\begin{equation}\label{eq:CAGO_scaled_configuration}
    \mathbf{Q}_b^{\mathrm{sc}}=\mathbf{q}_0\mathbf{1}_M^T+\frac{f_{b_0}}{f_b}\left(\mathbf{Q}_{b_0}^{\mathrm{anc}}-\mathbf{q}_0\mathbf{1}_M^T\right),
\end{equation}
where $\mathbf{1}_M=[1,\ldots,1]^T\in\mathbb{R}^{M\times 1}$ denotes the $M$-dimensional all-ones vector. The resulting scaled configurations are projected onto the feasible MA set and combined with the uniform and anchor configurations to form the initialization candidate set. Since the exact spatial scaling in Proposition~\ref{Proposition:1} requires identical normalized MPC power profiles and pairwise MPC phase differences across bands, the scaled configurations are used only as initialization candidates and are evaluated using the actual multi-band channels.

For each candidate user-band assignment and MA configuration, its feasibility with respect to the SINR and total transmit-power constraints is examined by solving the minimum-power problem, given by
\begin{subequations}\label{eq:CAGO_feasibility}
\begin{align}
    P_{\min}(\mathbf{A},\mathbf{Q})\triangleq\min_{\{\mathbf{w}_{k,b}\}}~~&\sum_{b=1}^{B}\sum_{k\in\mathcal{K}_b}\|\mathbf{w}_{k,b}\|^2\label{eq:CAGO_feasibility_obj}\\
    \mathrm{s.t.}~~&\gamma_{k,b}\geq\gamma_k^{\mathrm{th}},\quad k\in\mathcal{K}_b,\ \forall b.\label{eq:CAGO_feasibility_constraints}
\end{align}
\end{subequations}
For fixed $\mathbf{A}$ and $\mathbf{Q}$, the phase of each precoding vector can be adjusted such that the corresponding desired signal is real and nonnegative. Accordingly, the SINR constraints in \eqref{eq:CAGO_feasibility_constraints} can be equivalently transformed into second-order-cone constraints, and problem \eqref{eq:CAGO_feasibility} is a second-order cone program (SOCP). A candidate configuration is feasible if $P_{\min}(\mathbf{A},\mathbf{Q})\leq P_{\max}$, and the corresponding minimum-power solution is used to initialize the transmit precoders. Among all feasible MA candidates, the configuration achieving the largest sum rate after several precoding updates is selected as $\mathbf{Q}^{(0)}$.

\subsection{WMMSE-MM Continuous-Variable Optimization}
\label{Sec:CAGO_continuous}

\subsubsection{WMMSE-Based Precoding Update}
\label{Sec:CAGO_WMMSE}

For a fixed user-band assignment and MA configuration, the transmit
precoding is optimized based on the WMMSE transformation. The total received power and MSE of $U_k$ over its assigned band $b$ are respectively expressed as
\begin{equation}\label{eq:CAGO_total_received_power}
    T_{k,b}\triangleq\sum_{j\in\mathcal{K}_b}\left|\mathbf{h}_k^{(b)H}(\mathbf{Q})\mathbf{w}_{j,b}\right|^2+\sigma_b^2,
\end{equation}
and
\begin{equation}\label{eq:CAGO_MSE}
    e_{k,b}\triangleq|u_{k,b}|^2T_{k,b}-2\operatorname{Re}\left\{u_{k,b}^{\ast}\mathbf{h}_k^{(b)H}(\mathbf{Q})\mathbf{w}_{k,b}\right\}+1,
\end{equation}
where $u_{k,b}$ denotes the receive equalizer of $U_k$. The achievable rate in \eqref{eq:achievable_rate} is equivalently expressed in the WMMSE form as
\begin{equation}\label{eq:CAGO_rate_WMMSE}
    R_{k,b} =\max_{u_{k,b},\,\rho_{k,b}>0} \frac{W_b}{\ln 2}\left( \ln\rho_{k,b}-\rho_{k,b}e_{k,b} +1\right),
\end{equation}
where $\rho_{k,b}$ denotes the corresponding MSE weight. For given transmit precoders and MA configuration, the optimal equalizer and MSE weight are obtained as
\begin{equation}\label{eq:CAGO_equalizer_update}
    u_{k,b}^{\star}=\frac{\mathbf{h}_k^{(b)H}(\mathbf{Q})\mathbf{w}_{k,b}}{T_{k,b}},
\end{equation}
and
\begin{equation}\label{eq:CAGO_weight_update}
    \rho_{k,b}^{\star}=\frac{1}{e_{k,b}^{\mathrm{MMSE}}},
\end{equation}
respectively, where $e_{k,b}^{\mathrm{MMSE}}=\frac{1}{1+\gamma_{k,b}}$. Accordingly, for a fixed user-band assignment, the sum-rate maximization is equivalently transformed into the minimization of
\begin{equation}\label{eq:CAGO_WMMSE_objective}
    \Phi\triangleq\sum_{b=1}^{B}\sum_{k\in\mathcal{K}_b}\frac{W_b}{\ln 2}\left(\rho_{k,b}e_{k,b}-\ln\rho_{k,b}\right).
\end{equation}
For fixed $\{u_{k,b}\}$, $\{\rho_{k,b}\}$, and $\mathbf{Q}$, the transmit precoders are updated by solving
\begin{subequations}\label{eq:CAGO_precoding_subproblem}
\begin{align}
    \min_{\{\mathbf{w}_{k,b}\}}~~&\sum_{b=1}^{B}\sum_{k\in\mathcal{K}_b}\frac{W_b\rho_{k,b}}{\ln 2} e_{k,b}\label{eq:CAGO_precoding_obj}\\
    \mathrm{s.t.}~~&e_{k,b}\leq\frac{1}{1+\gamma_k^{\mathrm{th}}},\quad k\in\mathcal{K}_b,\ \forall b,\label{eq:CAGO_precoding_MSE_constraint}\\
    &\sum_{b=1}^{B}\sum_{k\in\mathcal{K}_b}\|\mathbf{w}_{k,b}\|^2\leq P_{\max}.\label{eq:CAGO_precoding_power_constraint}
\end{align}
\end{subequations}
Problem \eqref{eq:CAGO_precoding_subproblem} is a convex quadratically constrained quadratic program (QCQP) for fixed
$\{u_{k,b}\}$, $\{\rho_{k,b}\}$, and $\mathbf{Q}$. Since
$e_{k,b}^{\mathrm{MMSE}}\leq e_{k,b}$, the MSE constraint in
\eqref{eq:CAGO_precoding_MSE_constraint} guarantees the corresponding SINR
requirement. The equalizers, MSE weights, and transmit precoders are
successively updated until convergence.

\subsubsection{MM-Based MA Position Update}
\label{Sec:CAGO_MA}

For a fixed user-band assignment, transmit precoders, equalizers, and MSE weights, the common MA configuration is optimized based on the MM method. The position-dependent part of the WMMSE objective is expressed as
\begin{equation}\label{eq:CAGO_MA_objective}
F(\mathbf{Q})\triangleq\sum_{b=1}^{B}\sum_{k\in\mathcal{K}_b}\frac{W_b\rho_{k,b}}{\ln 2}e_{k,b}(\mathbf{Q}).
\end{equation}
Due to the wavelength-dependent spatial phase terms in the multi-band channels, both $F(\mathbf{Q})$ and the corresponding MSE constraints are non-convex with respect to the MA positions. Based on \eqref{eq:channel_coefficient}, the first- and second-order channel derivatives satisfy $\left\|\nabla_{\mathbf{q}_m}h_{k,m}^{(b)}(\mathbf{q}_m)\right\|\leq\frac{2\pi}{\lambda_b}\sqrt{\beta_k^{(b)}}\sum_{\ell=1}^{L_k}\sqrt{P_{k,\ell}^{(b)}}$,
and $\left\|\nabla_{\mathbf{q}_m}^{2} h_{k,m}^{(b)}(\mathbf{q}_m)\right\|\leq\left(\frac{2\pi}{\lambda_b}\right)^2\sqrt{\beta_k^{(b)}}\sum_{\ell=1}^{L_k}\sqrt{P_{k,\ell}^{(b)}}$.
These bounds characterize the different spatial variation rates across communication bands and provide band-dependent curvature information for constructing the MM surrogates. For $\psi\in\{F,e_{k,b}\}$, a quadratic majorizer around $\mathbf{Q}^{(t)}$ is constructed as
\begin{align}\label{eq:CAGO_MA_majorizer}
\widetilde{\psi}\left(\mathbf{Q}\mid\mathbf{Q}^{(t)}\right)=&\psi\left(\mathbf{Q}^{(t)}\right)+\left\langle \nabla\psi\left(\mathbf{Q}^{(t)}\right),\mathbf{Q}-\mathbf{Q}^{(t)}\right\rangle\notag\\
&+\frac{L_{\psi}^{(t)}}{2}\left\|\mathbf{Q}-\mathbf{Q}^{(t)}\right\|_F^2,
\end{align}
where $L_{\psi}^{(t)}>0$ is selected such that $\widetilde{\psi}(\mathbf{Q}\mid\mathbf{Q}^{(t)})\geq\psi(\mathbf{Q})$ within the trust region.

The non-convex minimum-spacing constraint is further handled through an inner approximation. For $\mathbf{d}_{i,j}^{(t)}\triangleq\mathbf{q}_i^{(t)}-\mathbf{q}_j^{(t)}$, the first-order lower bound of $\|\mathbf{q}_i-\mathbf{q}_j\|^2$ at the current MA configuration is given by $2\left(\mathbf{d}_{i,j}^{(t)}\right)^T(\mathbf{q}_i-\mathbf{q}_j)-\left\|\mathbf{d}_{i,j}^{(t)}\right\|^2\geq d_{\min}^2$.
Accordingly, the MA positions are updated by solving
\begin{subequations}\label{eq:CAGO_MA_subproblem}
\begin{align}
    \min_{\mathbf{Q}}~~&\widetilde{F}\left(\mathbf{Q}\mid\mathbf{Q}^{(t)}\right)\label{eq:CAGO_MA_subproblem_obj}\\
    \mathrm{s.t.}~~&\widetilde{e}_{k,b}\left(\mathbf{Q}\mid\mathbf{Q}^{(t)}\right)\leq\frac{1}{1+\gamma_k^{\mathrm{th}}},\quad k\in\mathcal{K}_b,\ \forall b,\label{eq:CAGO_MA_subproblem_MSE}\\
    &\mathbf{q}_m\in\mathcal{Q}, \quad\|\mathbf{q}_m-\mathbf{q}_m^{(t)}\|\leq\Delta^{(t)},\quad\forall m, \label{eq:CAGO_MA_subproblem_region}\\
    &2\left(\mathbf{d}_{i,j}^{(t)}\right)^T(\mathbf{q}_i\!-\!\mathbf{q}_j)\!-\!\left\|\mathbf{d}_{i,j}^{(t)}\right\|^2\!\geq\! d_{\min}^2,\quad \forall\, i<j. \label{eq:CAGO_MA_subproblem_spacing}
\end{align}
\end{subequations}
For a convex MA aperture region $\mathcal{Q}$, problem \eqref{eq:CAGO_MA_subproblem} is a convex QCQP. The curvature parameter $L_{\psi}^{(t)}$ is increased through backtracking whenever the corresponding majorization condition is not satisfied. Since the surrogate MSE constraints and the linearized spacing constraints constitute inner approximations of the original feasible set, each accepted MA positioning update remains feasible for problem P1.

\subsection{Analysis-Guided Assignment Refinement and Overall Algorithm}
\label{Sec:CAGO_assignment}

After the continuous-variable optimization converges, the user-band assignment is further refined through one-UE moves and pairwise swaps. For the current assignment, $b_k$ denotes the communication band assigned to $U_k$. A candidate move reassigns $U_k$ from $b_k$ to $b\in\mathcal{B}_k^{\mathrm{cand}}\setminus\{b_k\}$, and its analysis-guided score is defined as
\begin{equation}\label{eq:CAGO_move_score}
    \Delta S_{k,b}^{\mathrm{move}}\triangleq S_{k,b}^{\mathrm{UB}} -S_{k,b_k}^{\mathrm{UB}}.
\end{equation}
Similarly, the score associated with swapping the assigned bands of $U_k$ and $U_j$ is expressed as
\begin{equation}\label{eq:CAGO_swap_score}
    \Delta S_{k,j}^{\mathrm{swap}}\triangleq S_{k,b_j}^{\mathrm{UB}} + S_{j,b_k}^{\mathrm{UB}} -S_{k,b_k}^{\mathrm{UB}}-S_{j,b_j}^{\mathrm{UB}}.
\end{equation}
A pairwise swap is considered only when $b_j\in\mathcal{B}_k^{\mathrm{cand}}$ and $b_k\in\mathcal{B}_j^{\mathrm{cand}}$. For each candidate assignment $\mathbf{A}'$, the feasibility problem in \eqref{eq:CAGO_feasibility} is solved with the current MA configuration $\mathbf{Q}$. Infeasible candidates are discarded, while the corresponding minimum-power solution of each feasible candidate is used to initialize the WMMSE-based precoding optimization with $\mathbf{Q}$ fixed. The candidate achieving the largest sum rate is accepted only when
\begin{equation}\label{eq:CAGO_assignment_acceptance}
R_{\mathrm{sum}}\left(\mathbf{A}^{\mathrm{best}},\mathbf{Q},\mathbf{W}^{\mathrm{best}}\right)\geq R_{\mathrm{sum}}\left(\mathbf{A},\mathbf{Q},\mathbf{W}\right)+\epsilon_A.
\end{equation}
Therefore, the analytical scores in \eqref{eq:CAGO_move_score} and \eqref{eq:CAGO_swap_score} are used only to determine the evaluation order of the candidate assignments, while their feasibility and acceptance are determined using the original system constraints and actual achievable sum rate. The complete CAGO procedure is summarized in Algorithm~\ref{alg:CAGO}.
\begin{algorithm}[!t]
\caption{Cross-Band Analysis-Guided Optimization}
\label{alg:CAGO}
\begin{algorithmic}[1]
\Require Channel parameters, $P_{\max}$,
$\{\gamma_k^{\mathrm{th}}\}_{k=1}^{K}$, $\mathcal{Q}$, $d_{\min}$,
and thresholds $\epsilon_{\mathrm{W}}$, $\epsilon_{\mathrm{C}}$, and
$\epsilon_A$.
\Ensure $\mathbf{A}^{\star}$, $\mathbf{Q}^{\star}$, and
$\mathbf{W}^{\star}$.

\State Construct $\{\mathcal{B}_k^{\mathrm{cand}}\}$ using
\eqref{eq:CAGO_screening}; stop if any set is empty.

\State Solve \eqref{eq:sum_rate_UB} and compute
$S_{k,b}^{\mathrm{UB}}$ and $\widetilde{n}_b$ using
\eqref{eq:CAGO_assignment_score} and
\eqref{eq:CAGO_target_load}, respectively.

\State Obtain $\mathbf{A}^{(0)}$ from
\eqref{eq:CAGO_initial_assignment}.

\State Generate the MA candidates using
\eqref{eq:CAGO_scaled_configuration}; select a feasible
$\mathbf{Q}^{(0)}$ via \eqref{eq:CAGO_feasibility} and initialize
$\mathbf{W}^{(0)}$.

\State Set
$(\mathbf{A},\mathbf{Q},\mathbf{W})
\leftarrow
(\mathbf{A}^{(0)},\mathbf{Q}^{(0)},\mathbf{W}^{(0)})$.

\Repeat

    \Repeat
        \State Update $\{u_{k,b}\}$ using
        \eqref{eq:CAGO_equalizer_update} and
        $\{\rho_{k,b}\}$ using
        \eqref{eq:CAGO_weight_update}.

        \State Update $\mathbf{W}$ via
        \eqref{eq:CAGO_precoding_subproblem} until the relative
        WMMSE variation is no larger than
        $\epsilon_{\mathrm{W}}$.

        \State Update $\mathbf{Q}$ via
        \eqref{eq:CAGO_MA_subproblem}, with curvature backtracking
        when required.

    \Until{the relative sum-rate improvement is no larger than
    $\epsilon_{\mathrm{C}}$}

    \State Construct and rank the candidate moves and swaps using
    \eqref{eq:CAGO_move_score} and \eqref{eq:CAGO_swap_score}.

    \State Evaluate the highest-ranked candidates using
    \eqref{eq:CAGO_feasibility} and the fixed-$\mathbf{Q}$ WMMSE
    update; obtain
    $(\mathbf{A}^{\mathrm{best}},\mathbf{W}^{\mathrm{best}})$.

    \If{\eqref{eq:CAGO_assignment_acceptance} is satisfied}
        \State
        $(\mathbf{A},\mathbf{W})
        \leftarrow
        (\mathbf{A}^{\mathrm{best}},\mathbf{W}^{\mathrm{best}})$.
    \Else
        \State \textbf{break}.
    \EndIf

\Until{no improving assignment is accepted}

\State \Return
$(\mathbf{A}^{\star},\mathbf{Q}^{\star},\mathbf{W}^{\star})
\leftarrow
(\mathbf{A},\mathbf{Q},\mathbf{W})$.

\end{algorithmic}
\end{algorithm}

\subsection{Convergence and Computational Complexity}
\label{Sec:CAGO_convergence}

We characterize the convergence of the proposed algorithm in the following proposition.

\begin{Proposition}\label{Proposition:CAGO_convergence}
The sum-rate sequence generated by the proposed CAGO algorithm is non-decreasing and convergent.
\begin{IEEEproof}
For fixed $\mathbf{A}$ and $\mathbf{Q}$, the equalizer and MSE-weight updates attain their respective optimal solutions in the equivalent WMMSE formulation, while the precoding update minimizes the WMMSE objective subject to the original power constraint and the MSE-based SINR constraints. Therefore, the resulting updates do not decrease the achievable sum rate. For fixed user-band assignment, transmit precoders, equalizers, and MSE weights, the MM surrogate is tight at the current MA configuration and majorizes the original WMMSE objective. Moreover, the surrogate MSE and spacing constraints constitute inner approximations of the corresponding original constraints. Hence, solving \eqref{eq:CAGO_MA_subproblem} does not increase the WMMSE objective while preserving feasibility. The user-band assignment is updated only when the acceptance condition in \eqref{eq:CAGO_assignment_acceptance} is satisfied. Consequently, each accepted update preserves feasibility and the resulting sum-rate sequence is non-decreasing. Since the achievable sum rate is upper bounded by $\overline{R}_{\mathrm{sum}}$ in Proposition~\ref{Proposition:3}, the sequence converges.
\end{IEEEproof}
\end{Proposition}

Under the standard tightness and first-order consistency conditions of MM, every limit point of the continuous-variable iterations for a fixed user-band assignment is a stationary point of the corresponding WMMSE-MM problem. Furthermore, each accepted assignment update increases the sum rate by at least $\epsilon_A$, and hence only a finite number of assignment updates can be accepted. When the complete move-and-swap neighborhood is evaluated, the final assignment is locally optimal with respect to the considered neighborhood operations. For a truncated neighborhood, the same property holds with respect to the evaluated candidates. Nevertheless, global optimality is not guaranteed due to the mixed-integer non-convex nature of problem P1.

We next characterize the computational complexity of the proposed CAGO algorithm. The numbers of WMMSE, MA positioning, and outer assignment iterations are denoted by $I_{\mathrm{W}}$, $I_{\mathrm{Q}}$, and $I_{\mathrm{O}}$, respectively, while $I_{\mathrm{B}}$ denotes the average number of MM backtracking trials. The computational costs of solving the precoding QCQP, MA positioning QCQP, and feasibility SOCP are denoted by $\mathcal{C}_{\mathrm{W}}$, $\mathcal{C}_{\mathrm{Q}}$, and $\mathcal{C}_{\mathrm{F}}$, respectively. With $N_{\mathrm{A}}$ candidate assignments evaluated in each outer iteration, the overall computational complexity is expressed as
\begin{equation}\label{eq:CAGO_complexity}
\mathcal{O}\left(I_{\mathrm{O}}\left[I_{\mathrm{Q}}\left(I_{\mathrm{W}}\mathcal{C}_{\mathrm{W}}+I_{\mathrm{B}}\mathcal{C}_{\mathrm{Q}}\right)+N_{\mathrm{A}}\left(\mathcal{C}_{\mathrm{F}}+I_{\mathrm{W}}\mathcal{C}_{\mathrm{W}}\right)\right]\right).
\end{equation}
The precoding and feasibility problems involve approximately $2MK$ real optimization variables, while the MA positioning subproblem contains $2M$ real position variables and $M(M-1)/2$ minimum-spacing constraints. Before screening and ranking, the move-and-swap neighborhood contains at most $\sum_{k=1}^{K} \left(|\mathcal{B}_k^{\mathrm{cand}}|-1\right) +K(K-1)/2$ candidates. The analysis-guided screening and ranking reduce the number of evaluated candidates to $N_{\mathrm{A}}\ll|\mathcal{N}(\mathbf{A})|$, while the minimum-power solutions obtained from \eqref{eq:CAGO_feasibility} provide effective initialization for the subsequent WMMSE updates and further reduce the practical computational cost.

\subsection{Cross-Band Graph Representation}

To capture the coupling among user-band assignment, multiuser interference, and the common MA configuration, we construct a heterogeneous graph $\mathcal{G}=(\mathcal{V},\mathcal{E})$. Each node $v_{k,b}\in\mathcal{V}$ represents a candidate user-band pair with $b\in\mathcal{B}_k^{\mathrm{cand}}$, and its feature vector is expressed as
\begin{align}
\label{eq:GT_node_feature}
\mathbf{x}_{k,b}=\Big[&\mathbf{z}_{k,b}^{\mathrm{MPC}\,T},\log\beta_k^{(b)},\log\overline{g}_{k,b},\log(1+\overline{\gamma}_{k,b}),\nonumber\\
& W_b,\sigma_b^2,\gamma_k^{\mathrm{th}},S_{k,b}^{\mathrm{UB}}\Big]^T.
\end{align}
The MPC feature $\mathbf{z}_{k,b}^{\mathrm{MPC}}$ is obtained by applying a shared attention-pooling network to $\{P_{k,\ell}^{(b)},\varphi_{k,\ell}^{(b)}, \mathbf{u}_{k,\ell}\}_{\ell=1}^{L_k}$. Therefore, the node feature incorporates both the frequency-dependent multipath characteristics and the analytical quantities derived in Section~\ref{Sec:Anal}. Two types of edges are constructed to characterize the coupling among the candidate user-band pairs. The candidate-band nodes associated with the same UE are connected to capture their cross-band competition, with $\mathcal{E}_{\mathrm{CB}}=\left\{(v_{k,b},v_{k,b'}):b\neq b'\right\}$. 
The nodes associated with different UEs over the same communication band are connected to characterize their potential multiuser-interference coupling, with $\mathcal{E}_{\mathrm{IB}}=\left\{(v_{k,b},v_{j,b}): k\neq j\right\}$. 
In addition, a global node $v_{\mathrm{G}}$ is connected to all user-band nodes to aggregate the network-wide information associated with the common MA configuration.
    
\subsection{Relation-Aware Graph Transformer}

The feature vector of each user-band node is mapped into an initial hidden representation as $\mathbf{g}_{k,b}^{(0)}=\phi_{\mathrm{in}}(\mathbf{x}_{k,b})$, where $\phi_{\mathrm{in}}(\cdot)$ denotes a shared feature encoder. The three edge relations characterize different coupling relationships among the graph nodes and are therefore associated with relation-specific attention weights. For relation $r\in\{\mathrm{CB},\mathrm{IB},\mathrm{G}\}$, the attention coefficient between two connected nodes at the $n$-th graph-transformer layer is expressed as
\begin{equation}\label{eq:GT_attention} \alpha_{i,j,r}^{(n)}=\mathop{\operatorname{softmax}}\limits_{j\in\mathcal{N}_{i,r}}\left(\frac{\left(\mathbf{W}_{\mathrm{Q},r}^{(n)}\mathbf{g}_i^{(n)}\right)^T\left(\mathbf{W}_{\mathrm{K},r}^{(n)}\mathbf{g}_j^{(n)}\right)}{\sqrt{d_{\mathrm{h}}}}\right),
\end{equation}
where $\mathcal{N}_{i,r}$ denotes the neighbor set of node $i$ under relation $r$, and $d_{\mathrm{h}}$ denotes the hidden dimension. Accordingly, the relation-aware node representation is updated as
\begin{align}\label{eq:GT_update}
\mathbf{g}_i^{(n+1)}=\operatorname{LN}\Bigg(\mathbf{g}_i^{(n)}+\sum_r\sum_{j\in\mathcal{N}_{i,r}}\alpha_{i,j,r}^{(n)}\mathbf{W}_{\mathrm{V},r}^{(n)}\mathbf{g}_j^{(n)}\Bigg),
\end{align}
followed by a feed-forward network with a residual connection, where $\operatorname{LN}(\cdot)$ denotes layer normalization. The resulting user-band embeddings characterize both the cross-band assignment coupling and intra-band interference relationships, while the global embedding aggregates the network-wide information associated with the common MA configuration.

\subsection{Assignment and MA-Configuration Decoding}

The assignment probability associated with each candidate user-band pair $(k,b)$ is expressed as
\begin{equation}\label{eq:GT_assignment_probability}
    \pi_{k,b} =\frac{\exp\left(\phi_{\mathrm{A}}\left(\mathbf{g}_{k,b}^{(N_{\mathrm{GT}})}\right)\right)}{\displaystyle\sum_{j\in\mathcal{B}_k^{\mathrm{cand}}}\exp\left(\phi_{\mathrm{A}}\left(\mathbf{g}_{k,j}^{(N_{\mathrm{GT}})}\right)\right)},
\end{equation}
where $\phi_{\mathrm{A}}(\cdot)$ denotes the assignment decoder and $N_{\mathrm{GT}}$ is the number of relation-aware layers. Accordingly, the user-band assignment is determined as $a_{k,b}^{\mathrm{GT}}=\mathbbm{1}\big\{b=\arg\max_{j\in\mathcal{B}_k^{\mathrm{cand}}}\pi_{k,j}\big\}$. Since the assignment probabilities are normalized over $\mathcal{B}_k^{\mathrm{cand}}$, each UE is assigned to one candidate communication band, while the analytical screening in \eqref{eq:CAGO_screening} excludes user-band pairs that cannot satisfy the required SINR.

The common MA configuration is obtained from the global-node representation as $\widehat{\mathbf{Q}}=\mathbf{Q}^{\mathrm{ref}}+\operatorname{Reshape}\left(\phi_{\mathrm{Q}}\left(\mathbf{g}_{\mathrm{G}}^{(N_{\mathrm{GT}})}\right)\right)$,
where $\phi_{\mathrm{Q}}(\cdot)$ denotes the MA-configuration decoder and $\mathbf{Q}^{\mathrm{ref}}$ is a uniformly spaced feasible MA configuration. Since $\widehat{\mathbf{Q}}$ does not necessarily satisfy the aperture and minimum-spacing constraints, the feasible MA configuration is obtained by solving
\begin{subequations}\label{eq:GT_MA_restoration}
\begin{align}
    \mathbf{Q}^{\mathrm{GT}}&=\arg\min_{\mathbf{Q}}\|\mathbf{Q}-\widehat{\mathbf{Q}}\|_F^2\\
    \mathrm{s.t.}~~&\mathbf{q}_m\in\mathcal{Q},\quad \forall m,\\
    &2\left(\mathbf{d}_{i,j}^{\mathrm{ref}}\right)^T(\mathbf{q}_i\!-\!\mathbf{q}_j)\!-\!\left\| \mathbf{d}_{i,j}^{\mathrm{ref}}\right\|^2\geq d_{\min}^2,\quad \forall\, i<j,
\end{align}
\end{subequations}
where $\mathbf{d}_{i,j}^{\mathrm{ref}}\triangleq\mathbf{q}_i^{\mathrm{ref}}-\mathbf{q}_j^{\mathrm{ref}}$. For a convex MA aperture region $\mathcal{Q}$, problem \eqref{eq:GT_MA_restoration} is convex and always feasible since $\mathbf{Q}^{\mathrm{ref}}$ satisfies all its constraints. Moreover, the linearized minimum-spacing constraints constitute an inner approximation of the original spacing constraints. Therefore, the resulting $\mathbf{Q}^{\mathrm{GT}}$ satisfies \eqref{eq:P1_MA_region} and \eqref{eq:P1_MA_spacing}.

\subsection{Training and Online Refinement}
\vspace{-0.2em}
The CBGT model is trained offline using the solutions obtained from multi-start CAGO. For training realization $n$, $\mathbf{A}_{\mathrm{C}}^{(n)}$ and $\mathbf{Q}_{\mathrm{C}}^{(n)}$ denote the corresponding user-band assignment and MA configuration, respectively. The training loss is expressed as $\mathcal{L}_{\mathrm{GT}} =\lambda_{\mathrm{A}}\mathcal{L}_{\mathrm{A}}+\lambda_{\mathrm{Q}}\mathcal{L}_{\mathrm{Q}} -\lambda_{\mathrm{R}}\widetilde{R}_{\mathrm{sum}}$, 
where the assignment loss is defined as $\mathcal{L}_{\mathrm{A}}\!=\!-\!\sum_{n,k,b}a_{\mathrm{C},k,b}^{(n)}\log\pi_{k,b}^{(n)}$. To account for the permutation ambiguity among the MA indices, the MA-configuration loss is expressed as $\mathcal{L}_{\mathrm{Q}} =\sum_n\min_{\boldsymbol{\Pi}\in\mathcal{P}_M}\left\|\mathbf{Q}_{\mathrm{GT}}^{(n)} -\mathbf{Q}_{\mathrm{C}}^{(n)}\boldsymbol{\Pi}\right\|_F^2$, 
where $\mathcal{P}_M$ denotes the set of $M\times M$ permutation matrices. Moreover, $\widetilde{R}_{\mathrm{sum}}$ denotes a differentiable surrogate sum rate evaluated using the soft assignment probabilities and a fixed number of unfolded WMMSE iterations. The assignment and MA-configuration losses guide the CBGT model toward the CAGO solutions, while the surrogate sum-rate term further promotes the original sum-rate maximization objective.

For online implementation, the assignment probabilities in \eqref{eq:GT_assignment_probability} are used to construct a small set of high-probability candidate assignments. The feasibility of each candidate is examined through \eqref{eq:CAGO_feasibility} using $\mathbf{Q}^{\mathrm{GT}}$, and infeasible candidates are discarded. The feasible candidates are further evaluated after several WMMSE precoding updates, and the candidate achieving the largest sum rate is selected. Subsequent WMMSE-MM refinement jointly updates the transmit precoders and MA configuration. Consequently, the final solution satisfies the original assignment, SINR, transmit-power, aperture, and minimum-spacing constraints, while the graph-transformer prediction provides an effective initialization for reducing the required online optimization iterations.

\begin{table}[t]
\centering
\caption{Simulation parameters.}\vspace{-0.8em}
\label{tab:System_Para}
\renewcommand{\arraystretch}{1.05}
\setlength{\tabcolsep}{3.5pt}
\begin{tabular}{|l|c|c|}
\hline
\textbf{Parameter} & \textbf{Symbol} & \textbf{Value} \\ \hline
Number of UEs & $K$ & $6$ \\ \hline
Number of MAs & $M$ & $4$ \\ \hline
MA aperture region & $\mathcal{Q}$ & $0.6\times0.6$ m$^2$ \\ \hline
Minimum MA spacing & $d_{\min}$ & $0.058$ m \\ \hline
BS--UE distance & $d_k$ & $\mathcal{U}[80,100]$ m \\ \hline
Number of bands & $B$ & $3$ \\ \hline
Carrier frequencies & $\{f_b\}$ & $2.6,\,3.5,\,4.7$ GHz \\ \hline
Bandwidth & $W_b$ & $10$ MHz \\ \hline
Maximum transmit power & $P_{\max}$ & $5$ mW \\ \hline
SINR requirement & $\gamma_k^{\mathrm{th}}$ & $3$ dB \\ \hline
Number of MPCs & $L_k$ & $4$ \\ \hline
Path-loss exponent & $\eta$ & $2.5$ \\ \hline
Cluster shadowing std. & $\zeta$ & $6$ dB \\ \hline
Noise power & $\sigma_b^2$ & $-104$ dBm \\ \hline
\end{tabular}
\vspace{-1em}
\end{table}

The numbers of graph nodes and edges are denoted by $N_{\mathrm{V}}$ and $N_{\mathrm{E}}$, respectively. The inference complexity of the relation-aware graph transformer is $\mathcal{O}\left(N_{\mathrm{GT}}\left[N_{\mathrm{V}}d_{\mathrm{h}}^2+N_{\mathrm{E}}d_{\mathrm{h}}\right]\right)$.
Since the MPC encoder and graph-transformer parameters are shared across the UEs and communication bands, the proposed CBGT architecture is permutation equivariant with respect to the UE ordering. Additionally, the shared MPC encoder allows different numbers of MPCs to be processed without changing the trainable network parameters.

\section{Numerical Results}\label{Sec:Num}
In this section, we first present numerical results to validate the cross-band spatial scaling relationship established in Section~\ref{Sec:Anal}. We then evaluate the proposed CAGO and CBGT schemes and investigate the impacts of the number of UEs, number of MAs, and multipath richness, as well as the online computational cost. For each channel realization, the channel parameters are randomly generated according to the considered multi-band channel model and remain fixed during the corresponding optimization, while all schemes are evaluated under the same channel realizations and system parameters. For comparison, we consider five benchmark schemes: (i) the narrowband MA scheme selects the communication band achieving the largest optimized sum rate and jointly optimizes the MA positions and transmit precoders to serve all UEs over the selected band; (ii) the wideband MA scheme operates over a contiguous $30$~MHz band centered at $3.5$~GHz, divided into three $10$~MHz resource blocks, and jointly optimizes the user-resource assignment, common MA configuration, and resource-dependent transmit precoding; (iii) the fixed-antenna multi-band scheme employs uniformly spaced antennas and optimizes the user-band assignment and transmit precoding; (iv) the fixed-assignment multi-band scheme fixes the user-band assignment obtained from \eqref{eq:CAGO_initial_assignment} and optimizes the common MA configuration and transmit precoding; and (v) {the independent-MA multi-band scheme relaxes the common MA-configuration constraint by allowing each communication band to optimize a separate band-specific MA configuration, while retaining the user-band assignment and transmit precoding optimization. This physically idealized scheme provides an optimistic reference for assessing the performance loss caused by sharing a common MA configuration across bands.} All schemes employ the same number of antennas, total transmit-power budget, per-user SINR requirements, and underlying propagation geometry. The narrowband scheme uses a single $10$~MHz band, whereas all other schemes have a total available bandwidth of $30$~MHz. Unless otherwise specified, the simulation parameters are summarized in Table~\ref{tab:System_Para}.

\begin{figure}[t]
    \centering
    \subfigure[$f_1=2.6$ GHz, $\zeta=0$ dB.]{
        \includegraphics[width=0.4\columnwidth]{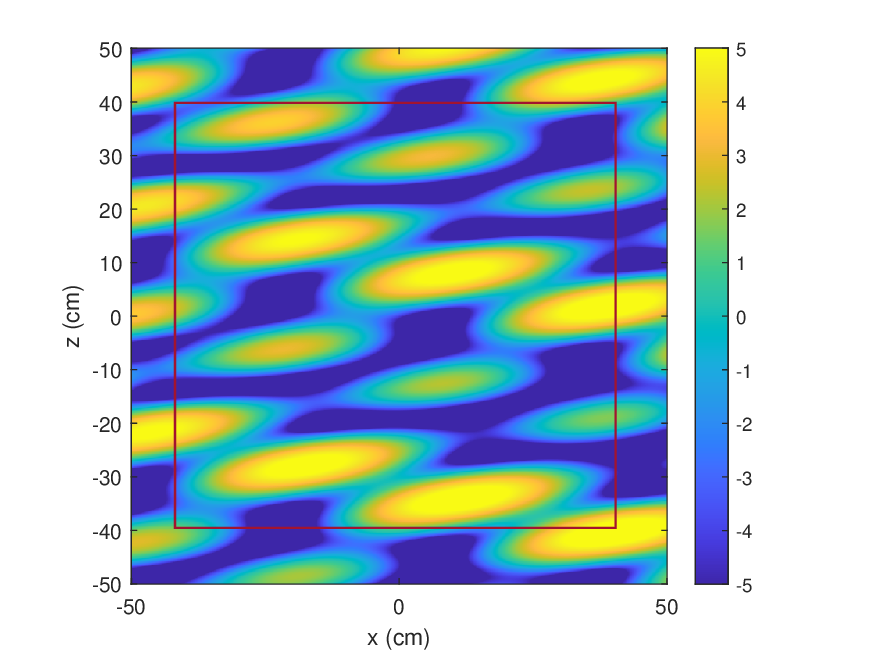}
        \label{fig:spatial_26_0}
    }\vspace{-1em}
    \subfigure[$f_3=4.7$ GHz, $\zeta=0$ dB.]{
        \includegraphics[width=0.4\columnwidth]{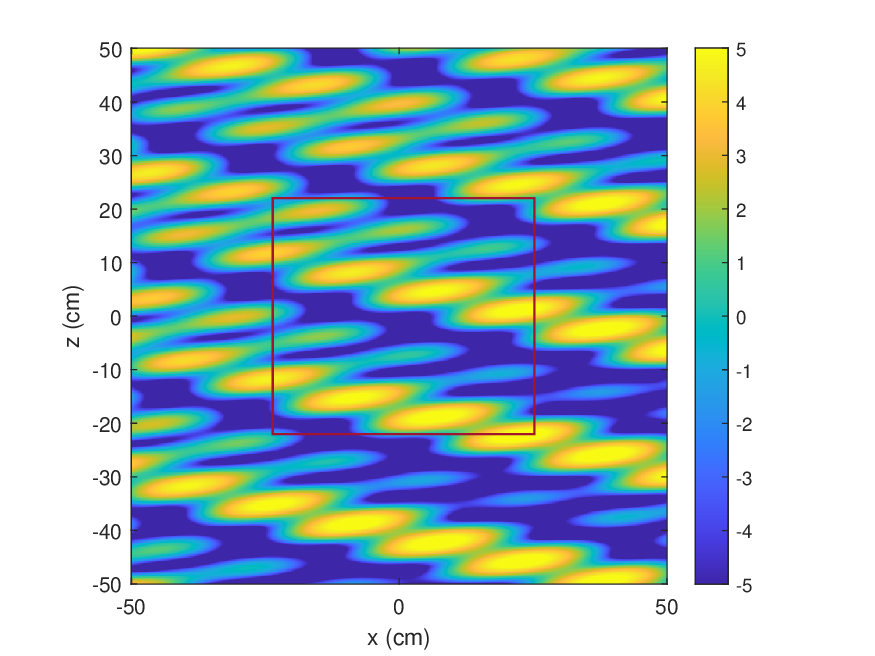}
        \label{fig:spatial_47_0}
    }
    \vspace{-1em}
    \subfigure[$f_1=2.6$ GHz, $\zeta=6$ dB.]{
        \includegraphics[width=0.4\columnwidth]{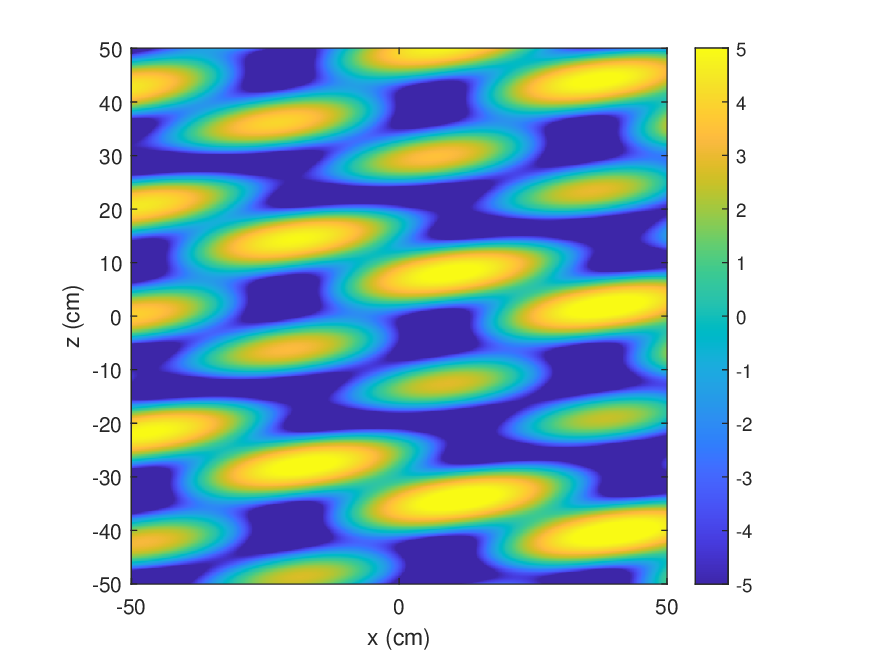}
        \label{fig:spatial_26_6}
    }
    \subfigure[$f_3=4.7$ GHz, $\zeta=6$ dB.]{
        \includegraphics[width=0.4\columnwidth]{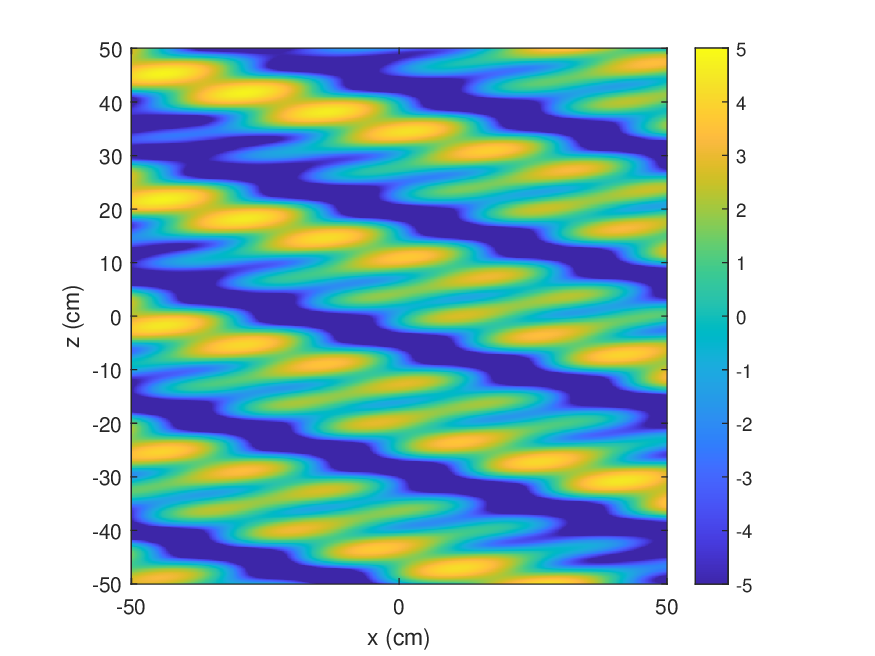}
        \label{fig:spatial_47_6}
    }
    \caption{Normalized channel power $G_k^{(b)}(\mathbf q)$ over two representative bands. Panels (a) and (b) consider $\zeta=0$ dB with identical normalized MPC power profiles and pairwise phase differences, while panels (c) and (d) consider $\zeta=6$ dB with band-dependent MPC powers and reference phases. The solid boxes indicate the regions under frequency-dependent spatial scaling.}
    \label{fig:cross_band_spatial_landscapes}
\end{figure}

Fig.~\ref{fig:cross_band_spatial_landscapes} plots the normalized channel power, $G_k^{(b)}(\mathbf q)$, over two representative communication bands. We first observe from Fig.~\ref{fig:cross_band_spatial_landscapes}(a) and (b) that the normalized channel power distributions within the solid boxes exhibit the same spatial structure under the corresponding frequency-dependent coordinate scaling, which verifies Proposition~\ref{Proposition:1}. Specifically, the corresponding spatial region at $4.7$ GHz is compressed relative to that at $2.6$ GHz according to the carrier-frequency ratio, while the normalized channel power pattern is preserved. For this validation case, we set $\zeta=0$ dB, resulting in identical normalized MPC power profiles across the two communication bands, and additionally set the pairwise MPC phase differences to be identical across bands. We then observe from Fig.~\ref{fig:cross_band_spatial_landscapes}(c) and (d) that the normalized channel power distributions exhibit different spatial structures and can no longer be related through simple coordinate scaling. This is because the normalized MPC powers and reference phases vary across the communication bands, although the underlying propagation geometry remains unchanged. Consequently, the favorable and unfavorable MA positions differ across bands, illustrating the cross-band MA configuration conflict induced by the frequency-dependent multipath characteristics.

\begin{figure}[t!]
    \centering
    \includegraphics[width=0.78\columnwidth]{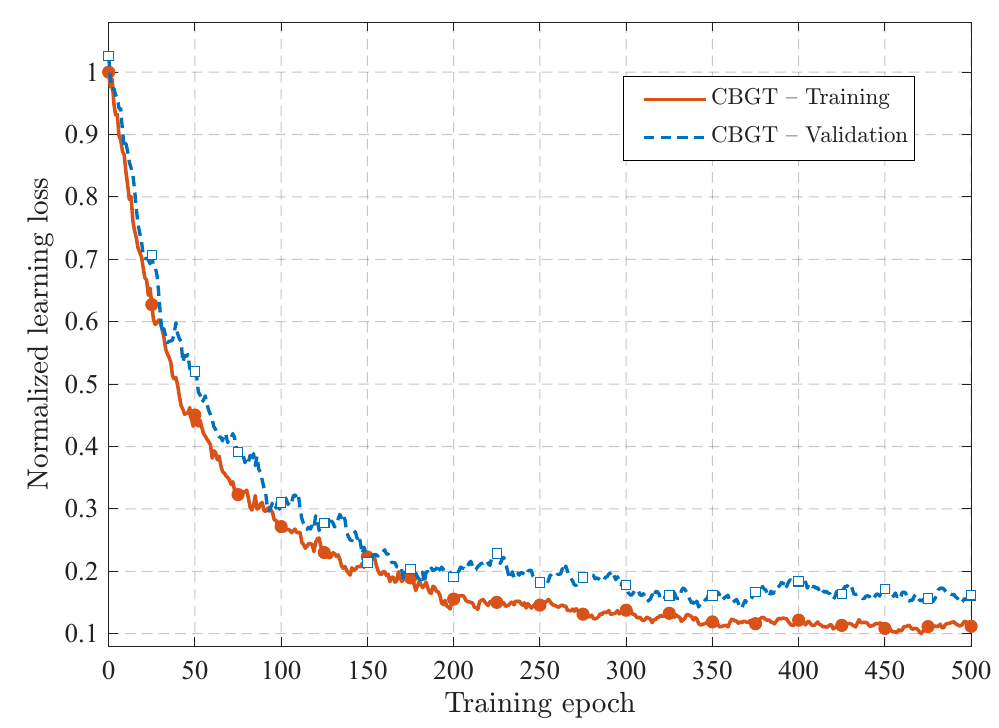}
    \vspace{-1em}
   \caption{Training and validation performance of the proposed CBGT in terms of the normalized learning loss.}\vspace{-1.5em}
\label{fig:CBGT_training}
\end{figure}

Fig.~\ref{fig:CBGT_training} shows the training and validation curves of the proposed CBGT in terms of the normalized learning loss, defined as the
weighted sum of the assignment and MA-configuration losses normalized by
its initial training-set value. We observe that both losses decrease rapidly during the early training stage and gradually converge as the number of epochs increases. The training loss becomes nearly stable after approximately $300$ epochs, while the validation loss follows a similar trend with moderate fluctuations. The final normalized training and validation losses are approximately $0.11$ and $0.16$, respectively, indicating stable training behavior with a relatively small generalization gap.

\begin{figure}[t!]
    \centering
    \includegraphics[width=0.78\columnwidth]{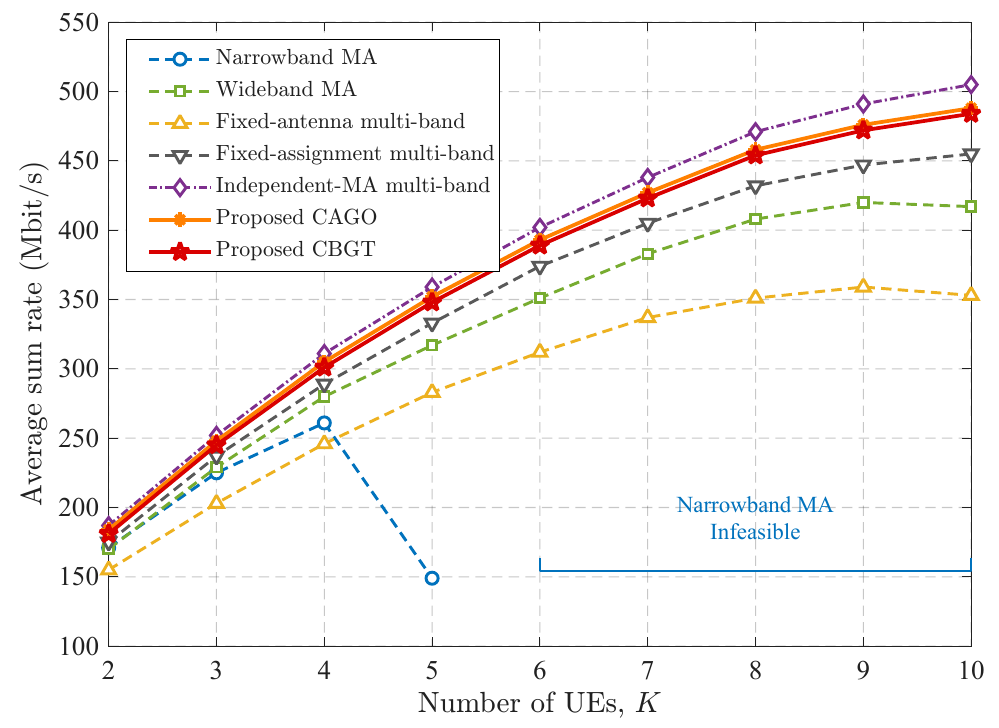}
    \vspace{-1em}
    \caption{Average sum rate versus the number of UEs, $K$.}
    \vspace{-0.5em}
    \label{fig:sr_vs_ue}
\end{figure}

Fig.~\ref{fig:sr_vs_ue} plots the average sum rate versus the number of UEs $K$ for $M=4$. We observe that the sum rates of CAGO, CBGT, and the independent-MA scheme generally increase with $K$, whereas those of the wideband and fixed-antenna schemes gradually saturate as $K$ becomes large. This observation reflects the tradeoff between the additional multiuser transmission opportunities and the increased difficulty of satisfying the per-user SINR requirements under a fixed transmit-power budget. The proposed CAGO scheme consistently outperforms the narrowband, wideband, fixed-antenna, and fixed-assignment schemes where feasible, while remaining close to the independent-MA benchmark. The gain over the fixed-antenna and fixed-assignment schemes demonstrates the benefits of adaptive MA positioning and adaptive user-band assignment, respectively, whereas the independent-MA scheme benefits from additional positioning flexibility across bands. Moreover, the proposed CBGT with online refinement achieves sum rates close to those of CAGO. In contrast, the sum rate of the narrowband scheme increases up to $K=4$ and decreases at $K=5$, while no feasible solution is obtained for $K\geq6$ under the considered constraints.

\begin{figure}[t!]
    \centering
    \includegraphics[width=0.78\columnwidth]{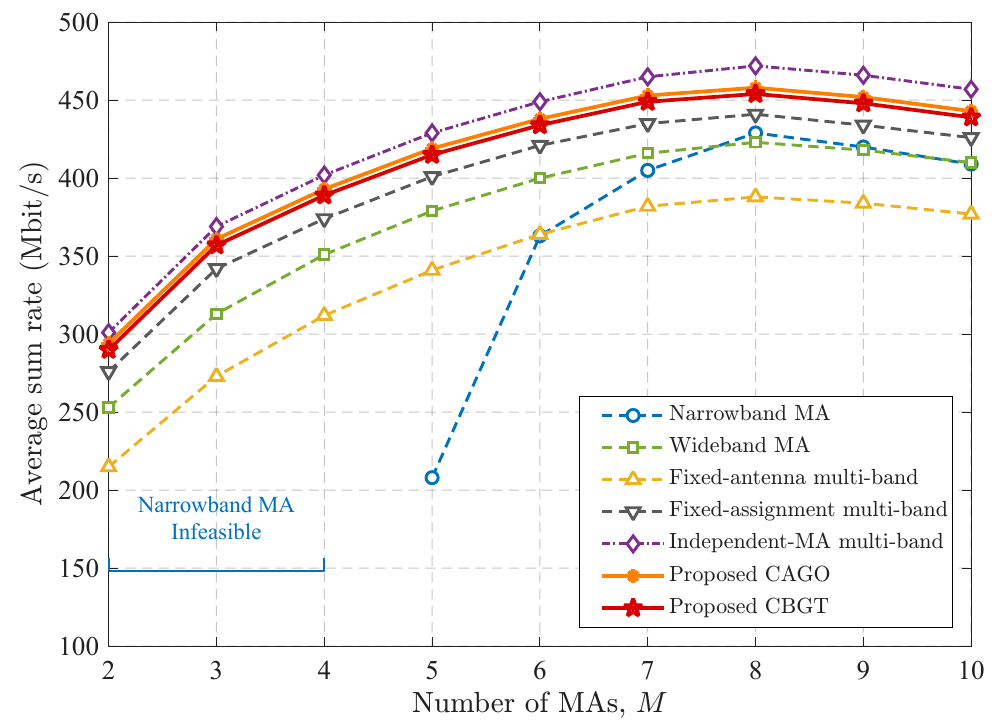}
    \vspace{-1em}
    \caption{Average sum rate versus the number of MAs, $M$.}\vspace{-0.5em}
    \label{fig:sr_vs_ma}
\end{figure}

Fig.~\ref{fig:sr_vs_ma} plots the average sum rate versus the number of MAs $M$ for $K=6$. We observe that the sum rates of all feasible schemes generally increase with $M$ up to around $M=8$, owing to the additional spatial DoFs that enhance the desired signals and facilitate multiuser interference suppression. The proposed CAGO scheme remains close to the independent-MA benchmark and consistently outperforms the other schemes over the considered range. The gains over the fixed-antenna and fixed-assignment schemes demonstrate the benefits of adaptive MA positioning and adaptive user-band assignment, respectively. Moreover, the proposed CBGT with online refinement achieves sum rates very close to those of CAGO. As $M$ further increases, the sum-rate improvement gradually diminishes and a slight decrease is observed for $M>8$. For the narrowband scheme, no feasible solution is obtained for $M\leq4$ under the considered constraints, while its sum rate improves significantly from $M=5$ to $M=6$, as the additional spatial DoFs facilitate the simultaneous transmission to all six UEs over a single band. The observed decline at larger $M$ does not establish an intrinsically optimal MA count. Whenever an existing configuration can be feasibly extended, assigning zero precoding weights to the additional MA preserves the original sum rate.

\begin{figure}[t!]
    \centering
    \includegraphics[width=0.78\columnwidth]{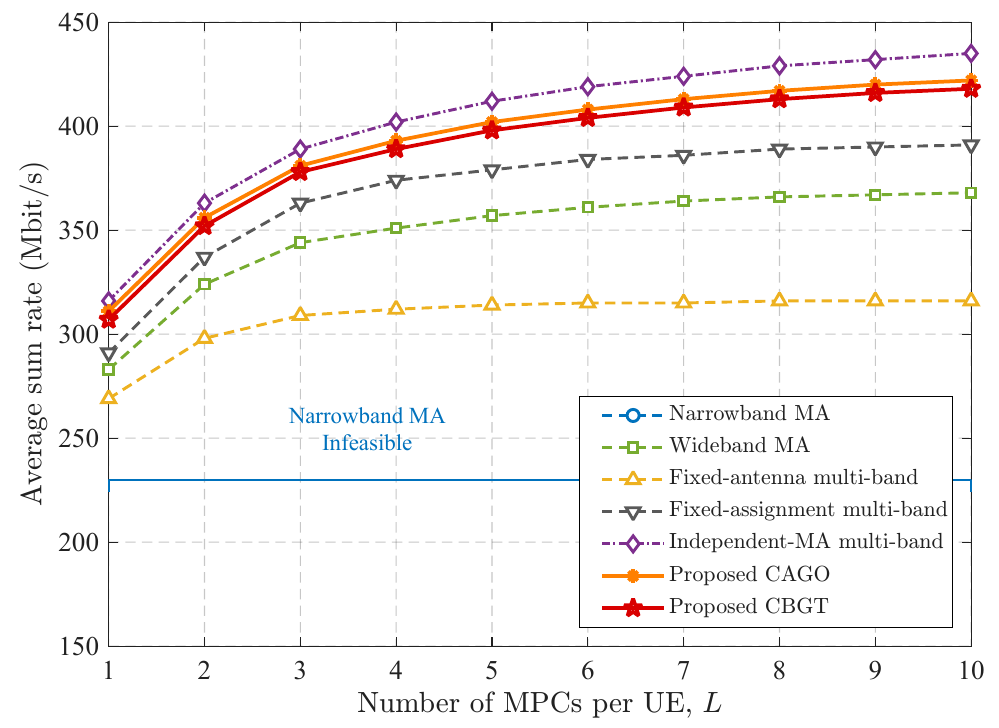}
    \vspace{-1em}
    \caption{Average sum rate versus the number of MPCs, $L$.}
    \vspace{-1em}
    \label{fig:sr_vs_L}
\end{figure}

Fig.~\ref{fig:sr_vs_L} plots the average sum rate versus the number of MPCs per UE, $L$. We first observe that the sum rates of all feasible schemes generally increase with $L$, while the corresponding gains gradually diminish as $L$ becomes large. This is because a richer multipath environment provides more diverse position-dependent spatial channel variations that can be exploited through MA positioning and user-band assignment, while the normalized total MPC power remains unchanged. We further observe that the proposed CAGO scheme remains close to the independent-MA benchmark as $L$ increases, while CBGT with online refinement achieves performance close to CAGO. In contrast, the wideband and fixed-antenna schemes exhibit more pronounced performance saturation for large $L$, indicating their more limited ability to exploit the increasingly diverse spatial channel variations. The narrowband scheme remains infeasible throughout the considered range under the specified SINR and transmit power constraints.

\begin{figure}[t!]
    \centering
    \includegraphics[width=0.78\columnwidth]{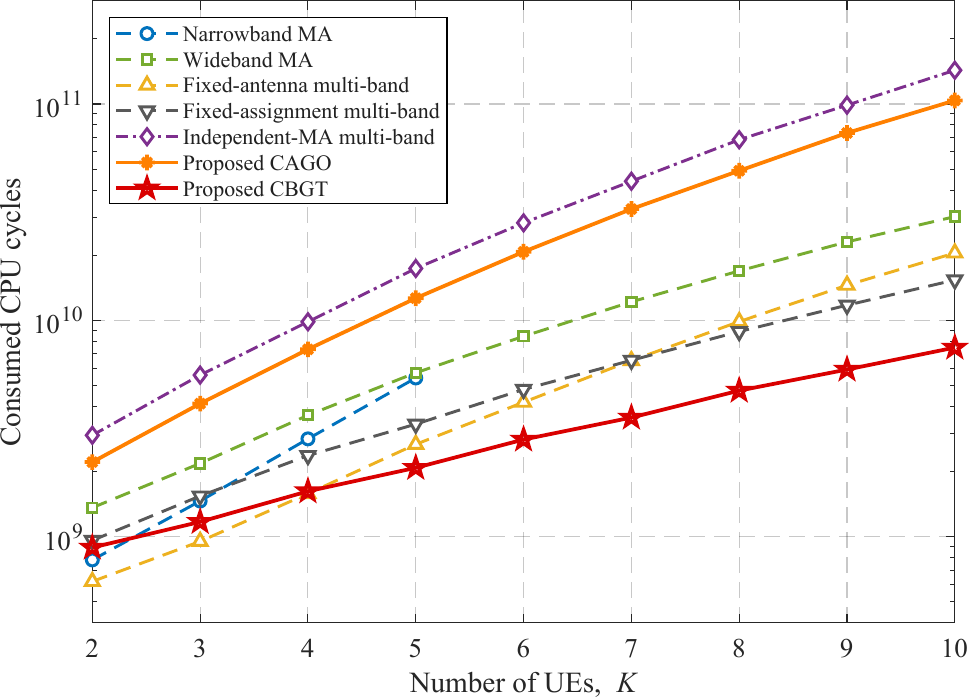}
    \vspace{-1em}
   \caption{Comparison of consumed CPU cycles versus the number of UEs, $K$.}\vspace{-0.5em}
\label{fig:CPU_cycles_vs_K}
\end{figure}

Fig.~\ref{fig:CPU_cycles_vs_K} compares the consumed CPU cycles versus the number of UEs, $K$. The computational cost increases with $K$ for all considered schemes, with the independent-MA scheme exhibiting the highest cost due to the separate MA optimization across communication bands. In contrast, the proposed CBGT requires substantially fewer CPU cycles than CAGO and consumes less than one tenth of its CPU cycles at $K=10$. The fixed-assignment and fixed-antenna schemes also incur lower computational costs than CAGO since they omit the assignment search and MA positioning optimization, respectively. Moreover, CBGT achieves the lowest computational cost among the feasible schemes for $K\geq5$. Together with its near-CAGO sum-rate performance observed in the previous results, these findings demonstrate that CBGT provides a favorable performance--complexity tradeoff and significantly reduces the online computational burden.

\section{Conclusion}\label{Sec:Con}
In this paper, we investigated the fundamental cross-band configuration problem in MA communications under frequency-dependent multipath propagation, where a common physical MA configuration is shared across distinct communication bands. We developed a multi-band channel model with shared propagation geometry and frequency-dependent MPC powers and phases, and jointly optimized user-band assignment, MA positioning, and transmit precoding. We established a cross-band spatial scaling relationship for the normalized channel power and derived position-independent channel-gain and sum-rate upper bounds, which characterize how frequency-dependent multipath variations reshape the spatial reconfiguration behavior of MAs across bands. Based on these analytical results, we developed the CAGO algorithm and further proposed the CBGT framework to reduce the online computational complexity. Numerical results validated the cross-band spatial analysis and showed that CAGO consistently outperforms the considered practical benchmarks while remaining close to the idealized independent-MA benchmark. They further revealed that CBGT with online refinement achieves near-CAGO sum-rate performance with substantially lower online computational cost, providing an effective approach for exploiting the spatial reconfiguration capability of MAs in practice.

\bibliographystyle{IEEEtran} 
\bibliography{bibli}

\end{document}